\documentclass[11pt,a4paper]{amsart}

\pdfoutput=1

\usepackage{natbib}

\usepackage{tikz}
\usepackage{mathrsfs}
\usepackage{amsfonts}

\usepackage{upgreek}

\DeclareFontFamily{OT1}{pzc}{}
\DeclareFontShape{OT1}{pzc}{m}{it}{<-> s * [1.10] pzcmi7t}{}
\DeclareMathAlphabet{\mathpzc}{OT1}{pzc}{m}{it}

\usepackage{geometry}                
\usepackage{graphicx}
\usepackage{amssymb}
\usepackage{epstopdf}
\usepackage{fullpage}
\usepackage{color}
\usepackage{float}
\usepackage{caption}
\usepackage{tablefootnote}

\usepackage{url}

\RequirePackage[OT1]{fontenc}
\RequirePackage{amsthm,amsmath}
\usepackage{amsmath, amssymb, amsfonts} 
\usepackage{algorithm}
\usepackage{algpseudocode}
\usepackage{booktabs}  
\usepackage{graphicx, subfigure} 
\usepackage{multirow} 

\usepackage{mathabx}

\usepackage{longtable}

\usepackage{accents}

\newcommand{\tildelow}[1]{%
  \accentset{\smash{\raisebox{-0.25ex}{$\scriptscriptstyle\sim$}}}{#1}%
}

\newcommand{\indep}{\ensuremath{\perp \!\!\! \perp}}

\newtheorem{definition}{Definition}[section]    

\begin{document}

\title[L\'evy Flight Cluster Model]{Modeling Human Spatial  Mobility Patterns with the L\'evy Flight Cluster Model}\thanks{CONTACT M. Wolff. Email: mlw32@uw.edu}

\author{Malcolm Wolff\textsuperscript{$\ast$}, Adrian Dobra\textsuperscript{$\ast$}, Anton H. Westveld \textsuperscript{$\dagger$} and Grace S. Chiu\textsuperscript{$\star\star$}}
\address{\textsuperscript{$\ast$} Department of Statistics, University of Washington, Seattle, WA, USA; \textsuperscript{$\dagger$} Research School of Finance, Actuarial Studies and Statistics, Australian National University, Australia; \textsuperscript{$\star\star$} William \& Mary's Batten School of Coastal \& Marine Sciences, Virginia Institute of Marine Science, VA, USA}

\begin{abstract}
Despite the extensive collection of individual mobility data over the past decade, fueled by the widespread use of GPS-enabled personal devices, the existing statistical literature on estimating human spatial mobility patterns from temporally irregular location data remains limited. In this paper, we introduce the L\'{e}vy Flight Cluster Model (\texttt{LFCM}), a hierarchical Bayesian mixture model designed to analyze an individual's activity distribution. The \texttt{LFCM} can be utilized to determine probabilistic overlaps between individuals' activity patterns and serves as an anonymization tool to generate synthetic location data. We present our methodology using real-world human location data, demonstrating its ability to accurately capture the key characteristics of human movement. 
\end{abstract}

\maketitle

\date{\today} 

\tableofcontents

\section{Introduction}

The collection of human activity data from the early twentieth century was based on travel journals \citep{wolf2001elimination, schonfelder2003activity, schonfelder2004variability}, surveys \citep{buliung2008exploring}, and social application queries \citep{luo2016explore}. 
Advancements in technology now include high-resolution GPS data \citep{hato2006development, stopher2008deducing, bohte2009deriving}, with smartphones offering a wealth of information that enables deeper analysis of human activity patterns \citep[e.g.,][]{kwan2012uncertain, matthews2013spatial, dobra2015spatiotemporal, williams2015measures}. 
For example, agent-based modeling (ABM) frameworks simulate autonomous actions to analyze behavioral shifts within larger patterns. This can reveal how individuals face geographical mobility constraints such as access to transportation \citep{kenyon2002transport} and epidemic-related stay-at-home orders \citep{perry2021pandemic}, linked to social exclusion and changes in employment opportunities \citep{goldhaber2015crossing}.

However, accurately estimating spatiotemporal trajectories from individuals' GPS data presents its own set of challenges. Human mobility displays a combination of both stochastic and decidedly deterministic movement patterns; localized activities like being at home, in an office, or at a gym, as well as traveling, lead to distinctive behavior patterns, each suggesting different degrees of social overlap--- probability of social interaction is high within a shared office space but practically nonexistent on a heavily trafficked highway. These behavioral patterns can also alter dramatically due to unobserved factors such as changes in financial status or the occurrence of a natural disaster. Since GPS data collected from mobile devices are primarily application dependent, the data consist of temporally irregular samples of individuals' time-stamped geolocations, which are closely related to their current activities. This introduces further complexity in inferring where and how the unrecorded time is spent, thus presenting an added challenge to the process of activity space estimation.

Early models of individual activity often relied on assumptions of multivariate normality, resulting in broad elliptical regions of probabilistic activity \citep{sherman2005suite, newsome1998urban}. Other proposed models defined activity regions as the convex hull around a specific set of points \citep[e.g.][]{buliung2006urban, fan2008urban}, which often unrealistically described the true activity area and encompassed large areas of uncertainty. More recently, methods such as kernel density estimation \citep[e.g.,][]{chen2020measuring, dong2020statistical} have been employed, which assume uniform distributions of activity within arbitrarily defined grid cells---where the estimation of activity overlap can be significantly influenced by the chosen cell size, as we show in Section \ref{sec:sim}. 

A separate body of research prominent in animal telemetry attempts to recover the latent position process that underlies the observed locations of individuals (see \citet{hooten2017animal} for an extensive review). By modeling the latent position process, they addressed the temporal dependence present in the observed data. Typically, these models rely on GPS data and employ Gaussian interpolation to deduce a latent position based on observed points and elapsed time \citep{hooten2017animal}. Over time, these models have seen various improvements, such as the inclusion of factors such as velocity, areas of attraction, and heterogeneous sampling times. For example, \citet{Scharf2017Imputation} introduce a hierarchical Bayesian multiple-imputation framework to fill gaps in sparse telemetry data, treating the true continuous path as a latent variable and demonstrating improved inference under measurement error. One notable model is the Brownian bridge movement model (BBMM) \citep{horne2007analyzing, kranstauber2012dynamic, kranstauber2014bivariate}, which determines individual latent positions using a sequence of Brownian bridges. However, BBMMs have been found to be inadequate in capturing the nuances of human mobility \citep{krumm2021brownian}; Brownian bridges do not allow ``loitering'' behavior or large movements for, for example, nonpedestrian travel, nor do they yield realistic inference when locations are detected not by GPS, but by a spatially static network of motion sensors that result in a highly irregular sampling rate that also varies between individuals \citep{ahmadian-et-2025}. Parallel to these advances, \citet{Scharf2015Dynamic} use discrete regime-switching indicators to infer dynamic social networks from movement data, illustrating the power of latent indicators for regime changes.

Characterizations of human mobility using non-Gaussian random-walk models have also been explored. For example, studies by \citet{gonzalez2008understanding} and \citet{rhee2011levy} have highlighted that human mobility often shows heavy-tailed travel distances similar to a Lévy flight process. \citet{song2010modelling} employed a modified Lévy flight process to model mobility, thus encapsulating geographic return patterns that are not adequately represented by random movement. In a similar vein, \citet{alessandretti2017multi} utilized logarithmic Gaussian mixtures in a discrete movement model to mirror the heavy-tailed nature of human mobility. Although these models represent characteristics of the latent process underlying observed individual mobility, they fail to provide uncertainty bounds for individuals' regions of highest activity based on local and global movement patterns. This gap underlines the remaining need for models capable of capturing the complex dynamics of human mobility while providing a realistic notion of uncertainty.

To address these gaps, we introduce the L\'{e}vy flight cluster model (\texttt{LFCM}) --- a generative Bayesian model for human mobility trajectories. Using the L\'{e}vy It\^{o} decomposition, we model individuals' movement as a L\'{e}vy flight exploration process that may exhibit a mixture of repetitive ``activity clusters" represented by local Brownian motions. Our model can be estimated from temporally irregularly sampled location data and with an arbitrarily small or large number of data points.
Unlike BBMMs, this model can represent non-movement (periods of time spent at the same location) as an individual's ``activity cluster", non-pedestrian movement as a L\'evy jump, and estimates propensity for an individual to return to each identified cluster.
Samples from the joint posterior distribution of the model's parameters can be used as a generative procedure for individuals' activity distributions, which are highly comparable to the individuals' original mobility patterns as measured by common mobility metrics.
This provides a novel approach to human mobility to data anonymization. By simulating the spatial mobility patterns of individuals with the \texttt{LFCM}, researchers can fit accurate models of human mobility based on the resulting synthetic data while preserving the privacy of the individuals involved.
Moreover, our generative model can be used an \textit{interpolative} and \textit{extrapolative} model, which allows one to use common metrics to compare distances between individuals even when individuals are not observed at the same times.
This means that, as we show in Section \ref{sec:appl}, one may generate probabilistic sociomatrices across any number of individuals, which can be used for later statistical inference.
Finally, our model can be efficiently estimated using analytical integration and Gibbs sampling methods and is completely parallelizable across individuals and trajectories.
For these reasons, \texttt{LFCM} contributes significantly to understanding human mobility, providing not only a nuanced model that captures the complex dynamics of human mobility but also establishing a foundation for further research to estimate activity areas and their overlap.

The structure of the paper is as follows. In Section \ref{sec:gps} we describe the characteristics of contemporary mobile device GPS data, specifically referring to GPS-generated geolocations collected at nonuniform time intervals from 293 anonymized devices within a metropolitan region for a duration of 12 weeks. In Section \ref{sec:metrics}, we examine commonly used accuracy metrics related to the study of individual mobility. These include jump length, radius of gyration, mean-squared displacement, and measures of frequented locations. In Section \ref{sec:review} we review previous research on individual mobility models. In Section \ref{sec:methods}
 we introduce the \texttt{LFCM} together with a Bayesian framework for inference and estimation of this model. In Sections \ref{sec:sim} and \ref{sec:appl} we evaluate the performance of the proposed model with simulated and real-world human location data. This assessment focuses on the model's ability to generate individual-level probabilistic activity areas, representative mobility sample paths for each individual, and accurate representation through the accuracy metrics mentioned above. To demonstrate an immediate practical application of the proposed model, in Section \ref{sec:appl} we illustrate the estimation of the probabilistic activity overlap between individuals. This investigation of the nuances of human mobility and its application in the definition of activity spaces serves to advance our understanding in the field and provides a solid foundation for future research on the relationships between human mobility and exogenous environmental factors.

\section{Mobile device GPS data} \label{sec:gps}

Mobile devices and their applications have become ubiquitous in the lives of people living in metropolitan areas. These devices and apps gather a wealth of data including timestamps, geographic location, communication content, and application-specific details, often under the purview of the relevant privacy policies. Data provisioning companies collect device data from mobile application developers or their aggregators. The types of data they accumulate include mobile ad identifiers (like Apple iOS IDFAs or Google Android IDs), time-stamped geographic coordinates of devices (latitude and longitude), the precision of these coordinates, and the device's direction and speed. 

\begin{figure}[!ht]
\begin{center}
    \includegraphics[scale=.35]{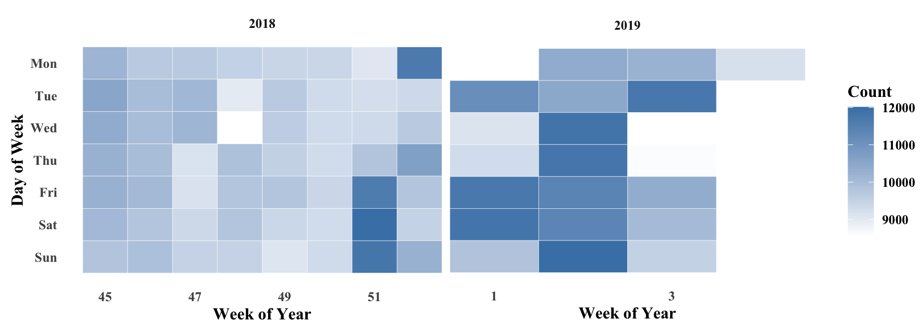}
    \caption{Distribution of records by day of week from November 2018 to January 2019.}
    \label{fig:time_of_year}
\end{center}
\end{figure}

In this article, we use the location records provided by a data provisioning company for 293 mobile devices active in a large metropolitan area. These data span a 12-week period from November 2018 to January 2019. Figure \ref{fig:time_of_year} displays the distribution of records by day of the week. Apart from the dates surrounding the winter holiday season, the records are relatively uniformly distributed across the period.

\begin{figure}[!ht]
    \begin{center}
        \includegraphics[scale=.62]{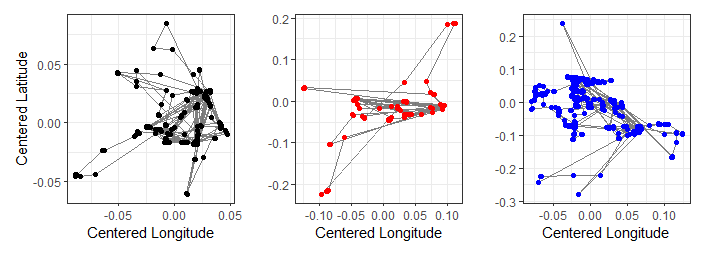}
    \end{center}
    \caption{Trajectory recorded for three individual devices over a one-week period starting on November 12, 2018. Latitude and longitude are shifted to maintain privacy.}
    \label{fig:example_path}
\end{figure}

Each record includes a unique device identifier, geographic coordinates, a timestamp, an estimated margin of error for the coordinates, and a reported speed of travel. This allows identification of daily movement patterns for individual devices over the 12-week period. Figure \ref{fig:example_path} presents the recorded trajectories of three different devices over a single week, starting November 12, 2018. It is evident from these trajectories that they contain both random and patterned components. They also highlight the intermittent temporal gaps in the recorded locations. Such gaps are common throughout this data set. 

\begin{figure}[htbp]
	\begin{center}
		\includegraphics[scale=.6]{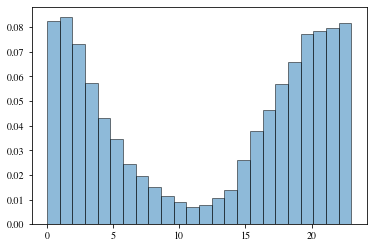}
	\caption{Distribution of the number of records by hour of day for an example individual over the 12 week period from 11/01/18 to 01/30/19.}
	\label{fig:hour_of_day}
	\end{center}
\end{figure}

Numerous mobile device applications carry out location-sensitive data collection while in operation. This leads to a significant fluctuation in timestamps by hour of the day and day of the year. Figure \ref{fig:hour_of_day} presents the average number of records by hour of the day for the same period (displayed at the bottom). Given this irregular rate of sampling, a temporally flexible method of estimating uncertainty around an individual's observable activity patterns becomes essential. The unique attributes of mobile GPS data, including its high resolution and irregular sampling rate, present challenges that require robust and flexible analytical methods to fully capitalize on potential of the data to capture human mobility trajectories.

\section{Human mobility metrics} \label{sec:metrics}

Quantification of individuals' daily living trajectories often uses four recognized metrics: jump length \citep{brockmann2006scaling, gonzalez2008understanding}, radius of gyration \citep{gonzalez2008understanding}, mean squared displacement \citep{alessandretti2017multi}, and frequent locations \citep{song2010modelling}. These metrics, applied within defined time periods, can capture unique aspects of human mobility. Many studies find that human travel often follows a heavy-tailed distribution ---- with many short trips and few long trips; jump length helps quantify this heterogeneity in travel behavior. The radius of gyration, as a measure of the typical distance an individual travels from their center of mass, is valuable for distinguishing between localized movement and more expansive travel patterns, indicating how confined or widespread an individual's activity space is. The mean squared displacement serves a similar purpose but measures displacement from the process center, not a single origin point. The number of frequented locations, assessed through a frequency rank, captures the regularity and predictability in routines and can help identify meaningful activity centers. Each metric serves as a reference point for the validity of the movement model in the context of individual mobility. Details for each of these metrics are given below.

\paragraph*{Jump Lengths} Let $\{\underline{X}(t) \in \mathbb{R}^2 : t \geq 0\}$ denote a spatio-temporal process. The \textit{jump length} of this process at time $t_i$ is the Euclidean distance between the locations $\underline{x}(t_{i-1})$ and $\underline{x}(t_i)$, denoted as 
\begin{equation}
    \begin{aligned}
        \Delta r(t_i) &\equiv  \|\underline{x}(t_i) - \underline{x}(t_{i-1})\|_2.
    \end{aligned}
\end{equation}
The jump length distribution $f_{\Delta r}(\Delta r(t))$ describes the likelihood that members of a population travel specific distances and is critical to accurately modeling population mobility. \citet{brockmann2006scaling} estimated the distribution of jump lengths for trajectories of U.S. bank notes according to the probability density function
\begin{equation}
    \begin{aligned}
        f_{\Delta r}(\Delta r(t)) &\propto [\Delta r(t)]^{-\alpha},
    \end{aligned}
\end{equation}
for the 95\% confidence interval $\alpha \in (1.57,1.61)$. This ``power law'' distribution has been consistently observed in further research on human mobility. \citet{gonzalez2008understanding} modeled jump lengths between individual GPS location data using a truncated version of this power law, 
\begin{equation}
    \begin{aligned}
        f_{\Delta r}(\Delta r(t)) &= [\Delta r(t) - \Delta r(t_0)]^{-\alpha}\exp(-\Delta r(t)/\kappa),
    \end{aligned}
    \label{eq:truncpowerlaw}
\end{equation}
finding a 95\% confidence interval for $\alpha \in (1.50, 1.90)$. Using two separate data sets on the location of GPS of individuals and the mode of transportation, \citet{zhao2015explaining} found that the aggregate distribution of $\Delta r(t)$ followed a power law distribution with $\alpha = 1.55$ and $\alpha =1.39$, respectively. 

\begin{figure}
    \centering
    \includegraphics[scale=.5]{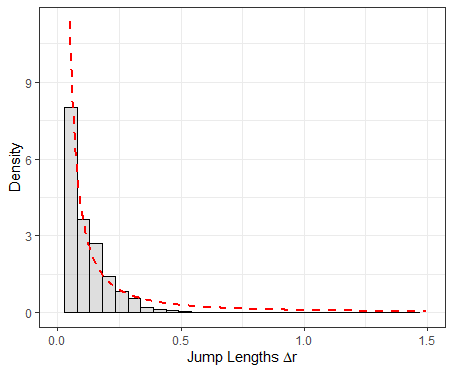}
    \caption{Histogram of jump lengths for 293 individuals in the analytic data set alongside the function $f_{\Delta r}(\Delta r) \propto (\Delta r)^{-1.59}$.}
    \label{fig:jump_length}
\end{figure}

Figure \ref{fig:jump_length} shows a histogram of jump length distributions for 293 individuals in the GPS mobility data described in Section \ref{sec:gps} alongside the density function $f_{\Delta r}(\Delta r(t)) \propto [\Delta r(t)]^{-1.59}$, where $1.59$ is the MLE for $\alpha$ found in \citet{brockmann2006scaling}, showing that the observed jump lengths roughly match a power-law distribution. 
However, this comparison assumes that jump lengths across individual/weeks are identically distributed. We also compute jump lengths between consecutive GPS locations within each (individual, week) group and analyze the upper tail of these lengths with a continuous power law model. For each group we select $x_{\min}$ by minimizing the Kolmogorov–Smirnov distance, then estimate the tail index $\alpha_g$ by maximum likelihood with asymptotic standard error $\mathrm{SE}(\hat\alpha_g)\approx(\hat\alpha_g-1)/\sqrt{n_{g,\text{tail}}}$. To obtain a single overall effect while accounting for between–group heterogeneity, we pool the $\hat\alpha_g$ via random–effects meta–analysis (REML), weighting by $w_g = 1/(\mathrm{SE}(\hat\alpha_g)^2+\tau^2)$, and find a 95\% CI for the pooled $\hat\alpha = (1.71, 1.74)$ with $\tau^2 = 0.075)$, suggesting jump lengths slightly above estimates in prior work.

\begin{figure}[htbp]
    \centering
    \includegraphics[scale=.32]{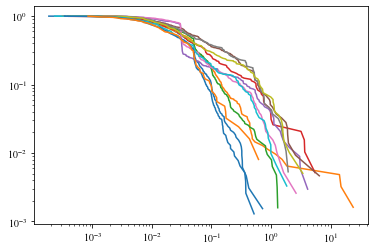}
    \includegraphics[scale=.32]{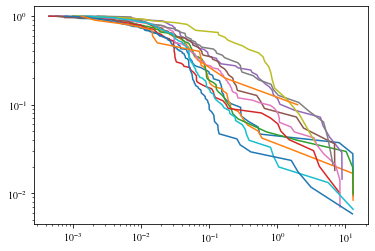}\includegraphics[scale=.32]{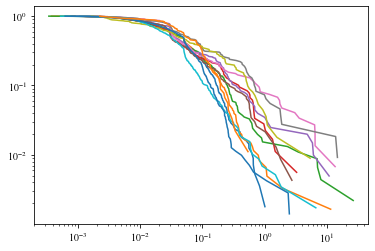}
    \caption{ECCDFs of jump distance on the log-log scale. Each plot is an individual device, and each curve is a week of observation.}
    \label{fig:ccdfs-by-week}
\end{figure}

Numerous papers have further corroborated the power law behavior of human mobility \citep[e.g.,][]{song2010modelling,rhee2011levy,alessandretti2017multi}. This behavior is evident at the individual level in our GPS mobility data as illustrated in  Figure \ref{fig:ccdfs-by-week} that shows the empirical complementary cumulative distribution functions (ECCDF) on the log-log scale. Each of the three plots represents a different individual device and each curve represents the trajectory over a week of observation. The approximate linearity of the tail in these distributions suggests heavy-tailed, or approximate power-law behavior.

\paragraph*{Mean Squared Displacement (MSD)} 
Characteristically, MSD quantifies the average squared distance an entity moves from its initial position up to a certain point in time; this can help infer the nature of the underlying movement process, e.g. whether it is diffusive or superdiffusive.

\begin{definition}[Mean Squared Displacement]
Let $\underline{x}(t_0)$ denote the initial point observed in the spatio-temporal process $\{\underline{X}(t) : t \geq 0\}$. The \textit{Mean-squared displacement} of the process up to time $t_n$ is defined by
\begin{equation}
    \begin{aligned}
        \text{MSD}(t_n) &:= \frac{1}{n}\sum_{i=1}^n \|\underline{x}(t_i) - \underline{x}(t_0)\|_2^2,
    \end{aligned}
\end{equation}
the average squared distance from the starting location $\underline{x}(t_0)$ up to time $t_n$. 
\end{definition}

As jump lengths are drawn from a heavy-tailed distribution, such as a power law or a truncated power law (see Equation~\eqref{eq:truncpowerlaw}), the process initially exhibits large fluctuations with occasional long jumps. However, truncation that is caused, for example, by travel restrictions limits the maximum jump length, ensuring that the distribution has a finite variance. Thus the central limit theorem guarantees that the sum of many independent jumps converges to a Gaussian distribution. Consequently, while short-time dynamics may reflect heavy-tailed behavior, long-term aggregated motion will asymptotically resemble a Brownian motion.

For the Brownian motion, the mean squared displacement (MSD) grows linearly with time:
\[
\langle r^2(t) \rangle \propto t,
\]
where the operator $\langle \cdot \rangle$ is often used to represent expectation in the statistical mechanics literature, which implies that the typical displacement (i.e., the square root of the MSD) scales as
\[
\sqrt{\langle r^2(t) \rangle} \propto t^{1/2}.
\]
Thus, despite the heavy-tailed nature of individual mobility events, the truncation forces the MSD to eventually exhibit the diffusive scaling characteristic of Brownian motion.

\paragraph*{Radius of Gyration.} The heavy-tailed behavior of humans on certain scales suggests that they tend to move a characteristic distance away from their starting locations. This distance is often quantified by the \textit{radius of gyration}, $r_g(t)$, defined as the root mean square distance of a set of points from a given axis. Traditional applications of this measure are rooted in physics (where it is related to the mass moment of inertia) and engineering (distribution of the cross-sectional area in a column). 
\begin{definition}[Radius of Gyration]
The Radius of Gyration $r_g(t)$ is given by:
\begin{equation}
    \begin{aligned}
    r_g(t_n) &= \sqrt{\frac{1}{n} \sum_{i=1}^n \|\underline{x}(t_i) - \overline{\underline{x}(t)}\|_2^2}.
    \end{aligned}
    \label{eq:rog}
\end{equation}
where $\underline{x}(t_i)$ are the $n$ geospatial coordinates of, e.g., an individual device trajectory, and $\overline{\underline{x}(t)}$ is mean of the trajectories $n$ points. 
\end{definition}
When applied to human mobility, the radius of gyration is often used to characterize the typical distance of an individual from the center of mass of their trajectory. \citet{gonzalez2008understanding} used mobile phone cell records to determine the radius of gyration for two sets of users; one set every two hours over one week and one set on a 6-month scale. The radius of gyration $r_g(t)$ was calculated for each user, and it was found that the distribution of $r_g(t)$ follows a truncated power law of the form

\begin{equation}
    \begin{aligned}
        f_{r_g}(r_g(t)) &= [r_g(t) - r_0]^{-\alpha_r}\exp{(-r_g(t)/\kappa_r)},
    \end{aligned}
\end{equation}
where $r_0$ represents a minimum radius of the gyration cut-off point due to the spatial sensitivity of the data, and $\kappa_r$ represents the upper cut-off due to the finite size of the study area. The measured 95\% confidence interval of the scaling exponent $\alpha_r \in (1.40,1.80)$---estimated from observations across both datasets---indicates a significant degree of heterogeneity in the travel habits of the observed population. Measuring the jump length distribution conditional on the radius of gyration $r_g$, $f_{\Delta r}(\Delta r(t) \ | r_g(t))$ revealed that users with small $r_g$ travel mostly over small distances, whereas those with large $r_g$ tend to display a combination of many small and a few larger jump sizes.

It is important to note that the distributions described above are not independent but instead are related by the equality 
$$f_{\Delta r}(\Delta r(t)) = \int_{r_0}^\infty  f_{\Delta r}\left(\Delta r(t) | r_g(t)\right)f_{r_g}\left(r_g(t)\right)d r_g(t),$$
where $f_{\Delta r}(\Delta r(t) | r_g(t))$ is the jump-length distribution introduced previously. If $f_{\Delta r}(\Delta r(t))$ has a power law scaling exponent $\alpha$, then we have $\alpha = \alpha_c + \alpha_r$ for conditional scaling exponent $\alpha_c$. This form of scaling corresponds to a type of random walk called a L\'evy flight, and the results suggest that this may be the behavior of individuals up to their associated characteristic distance $r_g$.

\paragraph*{Frequented Locations} A characteristic quality of human mobility is the frequent use of particular spaces or activity regions. Many people tend to travel to a specific region or along specific routes for work, return home on a daily basis, and frequent the same region to buy food, utilities, or clothing. In the literature on human mobility, it is common to quantify these frequent locations through the use of a classification system \citep{chen2020measuring}, and it has been shown that the probability of an individual returning to a particular location is inversely proportional to its frequent rank \citep{gonzalez2008understanding}. The tendency to return to known areas has also been quantified in the literature through the number of new locations visited over time, where GPS mobility data have been found to grow proportionally to $t^{0.6}$ \citep{song2010modelling}.

\section{Human mobility models} \label{sec:review}

We give an overview of the statistical frameworks used in the development of human mobility models. These models are treated as stochastic processes, subject to variability due to local movement and more extensive habitual travel. They are frequently modeled as structured random walks, a method that borrows concepts derived from estimation of animal mobility \citep[e.g.][]{liu2018temporal, meekan2017ecology}. These models also serve as the foundation for state-of-the-art mobility models. By examining these approaches, we can better understand the dynamics underlying individual-level mobility patterns.

\paragraph*{Random Walks} The concept of a random walk, first put forth by Karl Pearson in 1905 \citep{pearson1905problem}, is generally used to describe a sequence of random steps within a certain context, such as fluctuations in a gambler's wealth, the movement of a molecule, variations in a stock price, or the foraging behavior of an animal. A well-known model of a random walk is one based on a regular lattice, in which movement occurs on a discrete grid of locations according to a predetermined probability distribution. This lattice-based model can be expanded into a discrete random walk by considering movements across more complex spaces. Continuous-time random walks take this concept a step further by incorporating movements made within a continuous space over random time periods. A basic illustration of a continuous-time random walk within a two-dimensional real coordinate space ($\mathbb{R}^2$) can be represented by the following stochastic process
\begin{equation}
    \begin{aligned}
        \underline{X}(t) &= \underline{X}(t_0) + \sum_{i=1}^{n(t)} \underline{X}(t_i),
    \end{aligned}
\end{equation}
where $\underline{\Delta X}(t_i) \sim f_{\underline{\Delta X}}(\underline{\Delta X}(t))$ are the first differences and $n(t) \sim f_{n|t}(n|t)$ is a random number of jumps in the interval $(t_0, t)$. Then
\begin{equation}
    \begin{aligned}
        f_{\underline{X}(t)}(\underline{X}(t)) &= \sum_{n=0}^\infty f_{n| t}(n|t)\prod_{i=1}^{n(t)} f_{\underline{\Delta X}}(\underline{\Delta X}(t_i)).
    \end{aligned}
    \label{eq:ctrw}
\end{equation}
One property that plays an important role in the characterization of random-walk distributions is that of \textit{scale invariance}, wherein any magnification shows self-similarity to the whole. A function $f(x)$ is considered scale invariant under all rescalings if for any scalar $\lambda$,
\begin{equation}
    \begin{aligned}
        f(\lambda x) &= \lambda^{\delta}f(x)
    \end{aligned}
\end{equation}
for some exponent delta $\delta$. This property holds with $\delta = 0$ for white noise \citep{gardiner2009stochastic, vankampen2007stochastic}, and $\delta = -2$ for Brownian motion \citep{mandelbrot1968fractional, feller1971introduction, barabasi1995fractal}, for example.  

The property of scale invariance has been empirically seen in human mobility data through the use of power-law distributions to model jump lengths \citep[e.g.,][]{gonzalez2008understanding}. However, random walks alone do not adequately describe numerous other aspects of human mobility such as a nonuniform distribution of jump angles and nonlinear growth of newly visited locations.

\paragraph*{Brownian Motion} Brownian motion or the Wiener process is a stochastic process that is often used to model random movement in continuous time and space \citep{doob1953stochastic}. Although the original observations of Brownian motion were rooted in physical phenomena, its mathematical representation has broad applications, such as financial modeling or physics. 

\begin{definition}[Brownian motion]
A Brownian motion process $B(t)$ is characterized by the following properties:
\begin{itemize}
    \item[\textbf{1.}] $B(0) = 0$.
    \item[\textbf{2.}] Independent increments: for every $t > 0$, $$B(t + h) - B(t) \indep B(s),$$ for $u,s\geq 0$ and $s \leq t$.
    \item[\textbf{3.}] Gaussian increments: $B(t + h) - B(t) \sim \mathcal{N}(0, h)$.
    \item[\textbf{4.}] $B(t)$ is continuous in $t$.
\end{itemize}
\end{definition}

This process is also often described from a statistical mechanics perspective. For the probability density $\rho(x, t)$ of the particles at position $x$ and time $t$, Brownian motion satisfies the diffusion equation:
\begin{equation}
    \begin{aligned}
        \dfrac{\partial \rho(x,t)}{\partial t} &= D\dfrac{\partial^2 \rho(x,t)}{\partial x^2},
    \end{aligned}
    \label{eq:diffusion}
\end{equation}
where $D$ is the diffusion coefficient, a positive constant that quantifies the rate at which the probability mass spreads over time. If $n$ particles start at the origin at $t = 0$, Eq. \eqref{eq:diffusion} for $\sigma^2 = 2Dt$ has the solution $\rho(x, t) = \frac{n}{\sqrt{2\pi \sigma^2}}\exp(-\frac{1}{2\sigma^2}x^2)$, which is the density function of the normal distribution, suggesting that the mean square displacement of the Brownian motion is $\langle x^2 \rangle = 2Dt$. The form of the mean squared displacement in relation to time is often used to classify diffusion processes. For example, when the relationship between time and displacement is non-linear, $\langle x^2 \rangle \propto t^\gamma$, it is called an anomalous diffusion process, said to exhibit super-diffusion if $\gamma > 1$ and sub-diffusion when $\gamma < 1$. 

Brownian motion has a substantial history in describing movement processes in ecology. A seminal paper by \citet{horne2007analyzing} introduced the Brownian bridge movement model (BBMM) to study individual animal movement and regional occupation times. The model infers gaps in the observed process $\{\underline{X}(t_i) : i = 1,\ldots,n\}$ with two-dimensional Brownian bridges. That is, conditional on $\underline{x}(t_{i-1}) = \underline{a}$ and $\underline{x}(t_{i}) = \underline{b}$, $\underline{X}(t)$ for $t \in (t_{i-1}, t_i)$ is distributed as $\mathcal{N}\left(\underline{\mu}(t), \sigma^2(t)\mathbf{I}\right)$ where
\begin{equation}
    \begin{aligned}
        \underline{\mu}(t) \equiv \underline{a} + \frac{t - t_{i-1}}{t_i - t_{i-1}}(\underline{b}-\underline{a}) &&\text{and}&& \sigma^2(t) &\equiv \frac{(t - t_{i-1})(t_i - t)}{t_i - t_{i-1}}\sigma^2.
    \end{aligned}
    \label{eq:bbmm}
\end{equation}
The BBMM deals with serial correlation and unequal time intervals by directly incorporating this temporal independence.

\citet{kranstauber2012dynamic} extended the BBMM to incorporate heterogeneous activity in animal movement, calling the model the dynamic Brownian bridge movement model (dBBMM). The authors augment the original BBMM by assuming local dispersion parameters on subsets of the observed process calculated by sliding windows. By introducing the additional parameters window size $w$ and break point $b$, they are able to identify sudden changes in the animal's movement behavior. In separate work, \citet{kranstauber2014bivariate} develop another extension called the bivariate Gaussian bridge (BGB) model to allow variation in the dispersion parameter in two dimensions, one parallel and the other orthogonal to the angle between the start and end points. Together, these models are widely used in the literature on animal movement \citep[e.g.,][]{kays2015terrestrial, allen2016linking, rickbeil2019plasticity}. 

However, recent work by \citet{krumm2021brownian} has suggested that BBMMs do not accurately capture human mobility patterns. Considering dispersion parameters estimated by triplets of points no more than 50 kilometers apart and 48 hours apart, the author finds a large variation in diffusion coefficients specified at the group, individual, and triplet level, suggesting ambiguity in the level at which the human mobility model should be considered, and finds that intermittent points rarely satisfy the assumption of normality after applying the model. The author concludes that the dBBMM and BGB models may address some of these concerns; however, the author leaves analysis of these models to future work.

\paragraph*{L\'evy processes and flights} Brownian motion processes are a subset of a larger class of stochastic processes with stationary independent increments called L\'evy processes \citep{bertoin1996levy}. 

\begin{definition}[L\'evy process]
A stochastic process $X(t)$ is said to be a L\'evy process if it satisfies
\begin{itemize}
    \item[\textbf{1.}] $X(0) = 0$ almost surely.
    \item[\textbf{2.}] Independent increments: for any $0 \leq t_1 \leq t_2 \leq \dots t_n < \infty$, $$X({t_i}) - X(t_{i-1}) \indep X({t_j}) - X(t_{j-1}),$$ when $i \neq j$.
    \item[\textbf{3.}] Stationary increments: For $t > s$, $X(t) - X(s) \stackrel{d}{=} X(t-s)$.
    \item[\textbf{4.}] Continuity in probability: For any $\varepsilon > 0$ and $t \geq 0$,  $$\lim_{h \to 0} P(|X(t+h) - X(t)| > \varepsilon) = 0.$$
\end{itemize}
\end{definition}
Several key papers have found that several human mobility metrics show numerous statistical similarities with a particular type of L\'{e}vy process called a L\'evy flight. The term ``L\'evy flight'' was originally used by Beno\^it Mendelbrot \citep{cannon1984fractal} as a description of a heavy-tailed stochastic process $r(t)$ with step sizes defined by the survival function
\begin{equation}
    \begin{aligned}
        1 - F_r(r ; \varepsilon, \alpha) &= 
        \begin{cases}
            1 & r < \varepsilon,\\
            \left(\frac{\varepsilon}{r}\right)^\alpha & r \geq \varepsilon,
        \end{cases}
    \end{aligned}
    \label{eq:paretosurvival}
\end{equation}
commonly known as the Pareto distribution \citep{arnold1983pareto}. This distribution, which follows a power law according to its parameter $\alpha$, is widely used to model jump lengths in the human mobility literature. 

Although traditional L\'{e}vy flights are memoryless, human mobility exhibits repetitive returns to specific areas. Recent modeling techniques of human mobility attempt to explicitly incorporate this. \citet{song2010modelling}  develop an exploration and preferential return (EPR) model to match several intuitive aspects of human activity, modeling that the probability of visiting new locations decreases with time and a preference to return to a location with a probabilistic decision stage before moving: exploration with probability $P_{new} = \rho S^{-\gamma}$, or return with probability $P_{ret} = 1 - \rho S^{-\gamma}$. The conditional distance covered during an ``exploration'' is then modeled according to a heavy-tailed distribution $P(\Delta r)$, while during a ``return'' according to a distribution proportional to the number of previous visits. This alteration allows preservation of properties of L\'evy flight while agreeing on the propensity to visit new locations and the number of returns to particular locations.

\citet{alessandretti2017multi} offer an explanation for the apparent disconnect between the nonrandom intention of individuals and the scale-free nature of their movement, citing that mixtures of normal or log-normal distributions with different variances can generate a power law tail of movement often associated with scale-free distributions \citep{gheorghiu2004heterogeneity}. The authors model coordinate space as a nested structure of containers which humans traverse and movement as a two-step process in which the individual decides first which level of nested containers to move across and then which container to move in to. That is, the probability of transition from location $j$ to location $k$ is
$$p(j \to k) = p_{d(j,k)\mid d(j,h)}\prod_{l \leq d(j,k)} a(k_l),$$
where $p_{d(j,k)|d(j,h)}$ represents the probability of traveling at level distance $d(j,k)$ given a current level distance $d(j,h)$ from a home location $h$, and $\prod_{l \leq d(j,k)} a(k_l)$ is the probability of choosing a specific location $k$ at level distances $l \leq d(j,k)$. The size of each container, the maximum distance between any two points within a container, is described by a lognormal distribution, as is the time spent within each container. The authors find that this model is highly accurate in replicating certain characteristics of human mobility, such as the distribution of displacements, the radius of gyration, and visitation frequencies.

\section{The L\'{e}vy flight cluster model} \label{sec:methods}

Motivated by the literature gap for a generative model for the estimation of individual activity, we develop the L\'{e}vy flight cluster model (\texttt{LFCM}), a novel statistical framework to estimate individual activity distributions from a collection of few temporally irregular GPS location samples. The joint posterior distribution of the \texttt{LFCM}  parameters serve as a generative representation of individual activity distributions, which not only provides a novel approach to data anonymization, enabling researchers to simulate human mobility patterns while preserving the privacy of study participants, but also supports both interpolative and extrapolative analyses. As a result, comparisons of mobility between individuals are feasible even when data are collected at different times, allowing the creation of probabilistic sociomatrices for subsequent statistical inference. The remainder of this section develops the model progressively: we first lay the stochastic foundation via the L\'evy-It\^o decomposition and a jump-diffusion approximation; we then remove temporal dependence through a first-difference representation; next we introduce activity heterogeneity and memory via a finite mixture with returns; and we conclude with the Bayesian specification and priors for inference.

The \texttt{LFCM} leverages the L\'evy-It\^{o} decomposition theorem to model the latent stochastic process associated with an individual's trajectory as a memory-aware mixture of Brownian motions and pure jump processes. Through the use of conjugate priors \citep{ryan2017bayesian} almost all nuisance parameters can be integrated out, allowing the development of efficient posterior sampling methods.

\paragraph*{L\'evy-It\^{o} Decomposition} A fundamental result in the study of stochastic processes attributed to Paul L\'evy \citep{levy1938theorie} and Kiyosi It\^o \citep{ito1941stochastic} is the L\'evy-It\^{o} decomposition theorem, identifying a L\'evy process $\{X(t) : t\geq 0\}$ as the independent sum of a drifted Brownian motion process $\mu t + \sigma B(t)$, a compound Poisson process $J(t)$, and a square integrable martingale $\varsigma_\varepsilon(t)$ with almost surely countably many jumps with magnitude less than some specified $\varepsilon$. The compound Poisson process $J(t)$ is defined by 
\begin{equation*}
    \begin{aligned}
        J(t) &= \sum_{i=0}^{N(t)} \mathfrak{J}_i, \qquad t \geq 0,
    \end{aligned}
\end{equation*}
where $N(t)$ is a Poisson process and $\mathfrak{J}_i$ are independent random variables with distribution function $F_{\mathfrak{J}}$. Hence, the L\'evy process $X(t)$ can be written as
\begin{equation}
    \begin{aligned}
        X(t) &= \mu t + \sigma B(t) + J(t) + \varsigma_\varepsilon(t).
    \end{aligned}
    \label{eq:basicdecomp}
\end{equation}
If $\mathfrak{J}_i$ follows a symmetric distribution with power-law tails, decreasing as $|\mathfrak{J}|^{-\alpha - 1}$ for $0 < \alpha \leq 2$, the sum $J(t)$ will tend to a stable distribution \citep{gnedenko1954limit}; that is, for every $n\in N$, there exist constants $a_n > 0$ and $b_n \in \mathbb{R}$ such that
\begin{equation*}
    J_1(t) + J_2(t) + \cdots J_n(t) \overset{d}{=} a_nJ(t) + b_n,
\end{equation*}
where $J_1(t),J_2(t),\ldots,J_n(t)$ are i.i.d. copies of $J(t)$, and $\overset{d}{=}$ denotes equality in distribution.

\begin{figure}[htbp]
    \centering
    \includegraphics[scale=.55]{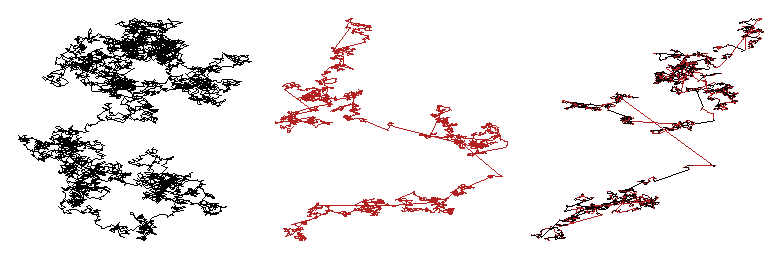}
    \caption{Three instances of random walks each run for 7,000 steps. Brownian motion steps are shown in black and L\'evy flight jumps are shown in red.}
    \label{fig:bm_lf_mix}
\end{figure}

\paragraph*{Jump Approximation} For stable L\'evy processes of index $\alpha \in (0,2]$, the martingale term $\varsigma(\varepsilon)$ is well approximated by Brownian motion in the sense that there exists $\sigma_\varepsilon$ such that the process $\sigma_\varepsilon^{-1}(\varsigma_\varepsilon(t) - \mathbb{E}[\varsigma_\varepsilon(t)]) \stackrel{d}{\rightarrow} B(t)$ as $\varepsilon \searrow 0$ \citep{asmussen2001approximations}. The L\'evy process in Eq. \eqref{eq:basicdecomp} then follows the approximate equality
\begin{equation}
    \begin{aligned}
     X(t) &\approx \mu t + (\sigma + \sigma_\varepsilon)B(t) + J(t)\\
     &\equiv \mu t + \sigma^* B(t) + J(t).
    \end{aligned}
    \label{eq:formulation}
\end{equation}
Figure \ref{fig:bm_lf_mix} illustrates the approximation using three instances of random walks in $\mathbb{R}^2$, each run for 7,000 steps. Brownian motion steps are shown in black and L\'evy flight jumps are shown in red; the left process is a Brownian motion, the middle process is a L\'evy flight, and the right process is a L\'evy flight with jumps less than $\varepsilon$ replaced by a Brownian Motion step. While the Brownian motion and L\'evy flight processes are distinct, with the latter exhibiting longer individual movements and the appearance of clustering, the right process in Figure \ref{fig:bm_lf_mix} has similar characteristics to a L\'evy flight.

\paragraph*{Temporal Independence} The temporal dependence in Eq. \eqref{eq:formulation} can be largely removed by considering the first difference representation
\begin{equation}
    \begin{aligned}
     \Delta X(t) &= \mu \Delta t + \sigma^* B(\Delta t) + J^*(\Delta t),
    \end{aligned}
    \label{eq:formulation2}
\end{equation}
since the components $B(\Delta t) \sim \mathcal{N}(\underline{0}, \Delta t \mathbf{I})$ are independent. Here, $\Delta X(t)$ can represent a change in geospatial location, $\Delta t$ a change in time, and the jump diffusion process $J^*(\Delta t)$ is either zero (i.e., $N(\Delta t) = 0$) or of magnitude strictly larger than some specified $\varepsilon$. Hence, Eq. \eqref{eq:formulation2} can be written as
\begin{equation}
    \begin{aligned}
     \Delta X(t) &= (\mu \Delta t + \sigma^* B(\Delta t))\mathbf{1}_{\{N(\Delta t) = 0\}} +  J^*(\Delta t)\mathbf{1}_{\{N(\Delta t) > 0\}}.
    \end{aligned}
    \label{eq:formulation3}
\end{equation}
Letting $\nu = \nu_{\Delta t} \equiv P(N(\Delta t) > 0)$, where we assume that the $N(\Delta t)$ are identically distributed, the process \eqref{eq:formulation3} has density

\begin{equation}
    \begin{aligned}
        f_{\Delta x}(\Delta X(t)\ |\ \mu, \sigma^*, \vartheta, \alpha, \Delta t) &= (1-\nu)\mathcal{N}(\underline{\Delta x}(t) ; \mu\Delta t, \sigma^*\Delta t) + \nu f_{J^*}(\Delta r(t) ;  \alpha, \Delta t),
    \end{aligned}
    \label{eq:simple_mixture}
\end{equation}

\par\noindent for a pure jump process distribution $f_{J^*}(\Delta r ;  \alpha)$ with jump length parameter $\alpha$. 

The above assumption highlights a key difference between L\'evy flights and L\'evy walks. In a Lévy flight, an individual (or object) makes instantaneous jumps between locations. These jumps can be of various lengths, but the probability of making a longer jump decreases as a power-law. In a Lévy walk, the individual (or object) still follows the power-law distribution for step lengths, but the motion within each step is not instantaneous. Instead, there is a defined speed, and longer steps will thus take more time to complete than shorter ones. Due to the sparsity of the data, and a lack of information on what the individual is doing, we make the assumption that the process is a L\'evy flight, rather than a L\'evy walk.

\paragraph*{Activity Heterogeneity} To account for heterogeneity in an individual's activities, the mixture formulation of $X(t)$ in Eq. \eqref{eq:simple_mixture} can be further extended to a finite mixture of $N_G$ processes:

\begin{equation}
    \begin{aligned}
     f_{\Delta x}(\Delta X(t) \mid N_G, \underline{\theta}, \Delta t) &\equiv \sum_{g=1}^{N_G}  \omega_g\biggl[(1-\nu)\mathcal{N}(\Delta x(t) ; \mu_g\Delta t, \sigma_g^*\Delta t)\\
     &\phantom{\equiv \sum_{g=1}^{N_G} \omega_g \biggl[ } + \nu f_{J^*}(\Delta r(t) ; \alpha_g, \Delta t)\biggr],
    \end{aligned}
    \label{eq:mixture_formulation}
\end{equation}

\par\noindent for mixture weights $\omega_g$ (e.g., the probability of a point belonging to activity $g$), where each $g^{th}$ mixture component is characterized by the tuple $(\mu_g, \sigma_g^*, \alpha_g)$, and $\underline{\theta}$ is the vectorized collection of model parameters for parsimony. Eq. \eqref{eq:mixture_formulation} allows the model to fit individuals with both a relatively low number of activities (e.g., living a sedentary lifestyle) and those with more active lifestyles. Eq. \eqref{eq:mixture_formulation} is easily adapted to a multivariate framework described by
\begin{equation}
    \begin{aligned}
        f_{\underline{\Delta x}}\left(\underline{\Delta X}(t) \mid N_G, \underline{\theta}, \Delta t \right) &\equiv \sum_{g=1}^{N_G} \omega_g\bigl[ (1-\nu) \mathcal{N}\left(\underline{\Delta x}(t) ; \underline{\mu}_g \Delta t, \mathbf{\Sigma}_g^*\Delta t \right)\\
        &\phantom{\equiv \sum \omega_g\ \biggl[}  +\nu f_{ J^*}(\underline{\Delta r}(t) ;\vartheta_g, \alpha_g, \Delta t)\bigr],
    \end{aligned}
    \label{eq:lfcm}
\end{equation}
where $N_G$ is the number of activity groups, $\underline{\mu}_g$ is a group-specific drift, $\mathbf{\Sigma}_g^*$ is the group-specific covariance matrix, and $f_{\Delta J^*}(\underline{\Delta r}(t) ; \vartheta_g, \alpha_g, \Delta t)$ is the distribution function of the pure jump process with jump angle parameter $\vartheta_g$ and jump length parameter $\alpha_g$. We assume the jump length and jump angle are independent, i.e. of the form $f_{J^*}(\underline{\Delta r}(t); \vartheta_g, \alpha_g, \Delta t) = \prod_{g=1}^{N_G} f_{\theta}(\theta ; \vartheta_g)f_{\Delta r}(\Delta r ; \alpha_g, \Delta t)$.

\begin{figure}[htbp]
    \centering
    \includegraphics[scale=.6]{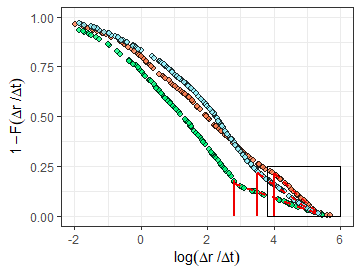}
    \includegraphics[scale=.6]{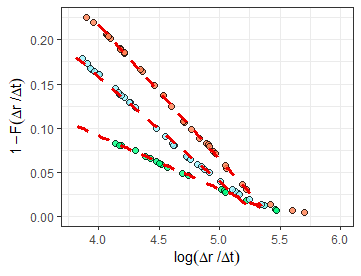}
    \caption{Survival functions for normalized jump lengths $\Delta r_t / \Delta t$ for three example individuals. The left panel depicts the entire distribution, whereas the panel figure depicts the zoomed in area denoted by the black box.}
    \label{fig:survival}
\end{figure}

\paragraph*{Pure Jump Size Distribution} Due to the complexity of aggregating a Poisson distributed number of arbitrary i.i.d. random variables, the compound Poisson process $J(t)$ is difficult to model in practice. Instead, we use a data-driven simplification of the density $f_{\Delta r / \Delta t}(\Delta r / \Delta t)$. A L\'{e}vy flight is characterized by the power-law tail of its step length, and the step lengths of human mobility as represented by GPS device location data observe a power-law tail distribution; Figure \ref{fig:survival} depicts complementary cumulative distribution functions (CCDF) for three distinct individuals observed in the data, each of which presents a linear tail on the log scale. This follows because
\begin{equation*}
    \nonumber
    \begin{aligned}
        1 - F_{\Delta r / \Delta t}(\Delta r / \Delta t) &\propto (\Delta r / \Delta t)^{-\alpha} \implies \log (1 - F_{\Delta r / \Delta t}(\Delta r / \Delta t)) \propto -\alpha \log(\Delta r / \Delta t). 
    \end{aligned}
\end{equation*}

\par\noindent We model jump lengths as a Pareto distribution with common parameters $\alpha, \varepsilon$ between activity groups, that is, 
\begin{equation*}
    \begin{aligned}
    f_{\Delta r / \Delta t}(\Delta r / \Delta t; \alpha, \varepsilon) &= \alpha \varepsilon^{\alpha} (\Delta r  / \Delta t)^{-(\alpha+1)}\mathbf{1}_{\{(\Delta r / \Delta t) \geq \varepsilon\}}.
    \end{aligned}
\end{equation*}

In specifying the Pareto distribution as a conjugate prior, we must provide an estimate of $\varepsilon$. Since the estimation of the parameter $\varepsilon$ involves a domain search, traditional MLE methods for joint estimation do not apply. Naive grid searches of the entire algorithm may be prohibitively expensive. Although one could still specify a prior distribution on $\varepsilon$, this removes the benefit of integrating the closed-form expression for the Pareto posterior, speeding up computational efficiency.\\

\paragraph*{Empirical Estimation of $\varepsilon$} We consider three separate methods of empirical estimation for $\varepsilon$---using the Kolmogorov-Smirnov (KS) statistic, the Anderson-Darling (AD) statistic, and the Kuiper's (K) statistic. The KS statistic is the maximum absolute difference between the empirical cumulative distribution function (ECDF) of a random variable $\varepsilon$, $\widehat{F}_n(\varepsilon)$, and a corresponding fitted parameterized distribution $\widehat{F}_{\varepsilon}(\varepsilon ; \theta)$:
\begin{equation*}
    \begin{aligned}
        \text{KS}(\varepsilon) &= \max_{x_i \geq \varepsilon} \left| \widehat{F}_n(x_i) - \widehat{F}_X(x_i) \right|.
    \end{aligned}
\end{equation*}
As an alternative to the KS statistic, the AD statistic is defined as
\begin{equation*}
    \begin{aligned}
        \text{AD}(\varepsilon) &= -n - \sum_{i=1}^n \dfrac{2i-1}{n}\left(\log \widehat{F}_\varepsilon(x_i) + \log \widehat{F}_\varepsilon(x_{n+1-i})\right).
    \end{aligned}
\end{equation*}
The AD statistic tends to be more sensitive to the tails of the distribution \citep{engmann2011comparing}. The KS statistic takes the pointwise maximum absolute difference of the ECDF and a fitted parameterized CDF, where this difference is unlikely to be largest at the tails. However, very small values of $\widehat{F}_{\varepsilon}(x_i)$ have a large weight in the AD statistic.

Kuiper's statistic (K) considers both positive and negative differences between the distributions. This makes the statistic equally sensitive to the tails as at the median and also makes it invariant under cyclic transformations of the independent variable. The statistic is defined by
\begin{equation*}
    \begin{aligned}
        K(\varepsilon) &= \max\left(\frac{i}{n} - \widehat{F}_\varepsilon(\varepsilon_i)\right) + \max\left(\widehat{F}_\varepsilon(\varepsilon_i) - \frac{i-1}{n}\right).
    \end{aligned}
\end{equation*}
Other formulations have been used, such as the Bayesian information criterion (BIC), but many of these do not perform well in low sample regimes \citep{clauset2009power}. 

Next we propose a new methodology for estimating $\varepsilon$ for small samples from power law generative distributions--- a neighborhood averaging procedure that improves estimates in low sample sizes without degrading estimates for larger samples. Let $\mathcal{E}$ be a range of potential values of $\varepsilon$. For a ball $B_{r}({\varepsilon})$ of radius $r$ around ${\varepsilon}$, let the $k$-smoothed neighborhood of $\varepsilon$ be defined as
\begin{equation*}
    \begin{aligned}
        \mathcal{E}_{\mathcal{N}} &\equiv \underset{\mathcal{N} \subset B_r(\varepsilon),\ |\mathcal{N}| = k}{\text{argmin}}\quad  \sum_{\varepsilon\in\mathcal{N}} D(\varepsilon),\\
    \end{aligned}
\end{equation*}
with corresponding estimates $\widehat{\varepsilon}^{\mathcal{N}} = \tfrac{1}{|\mathcal{E}_{\mathcal{N}}|} \Sigma_{\varepsilon\in\mathcal{E}_{\mathcal{N}}}\  \varepsilon$ and $\widehat{\alpha}_{\text{MLE}}^{\mathcal{N}} = \tfrac{1}{|\mathcal{E}_{\mathcal{N}}|}\Sigma_{\varepsilon\in\mathcal{E}_{\mathcal{N}}}\  \widehat{\alpha}_{\text{MLE}}(\varepsilon)$.

\begin{figure}[htbp]
    \centering
    \includegraphics[scale=.35]{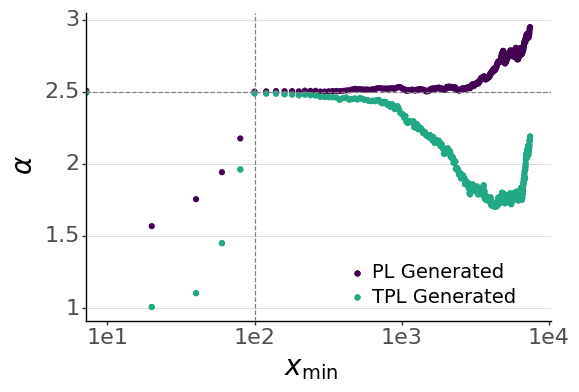}\includegraphics[scale=.35]{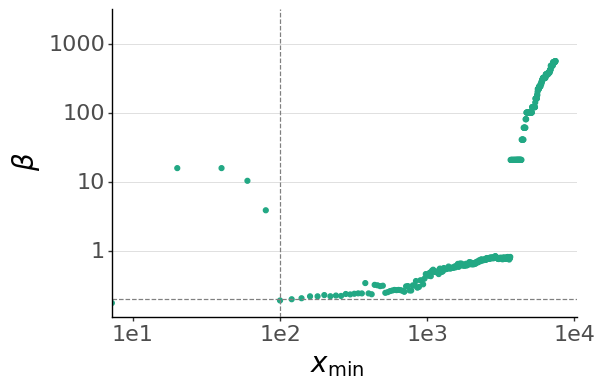}
    \caption{Mean estimate of $\alpha$ and $\beta$ for $5,000$ samples drawn from a Pareto distribution, with $\alpha = 2.5$, $\beta=0.2$, $\varepsilon = 100$, and $n = 2,500$, plotted as a function for $\varepsilon$. Estimates are generated with power law and truncated power law distributions.
    }
    \label{fig:clauset_tpl_pl}
\end{figure}

To show the implication of estimation bias when computing $\widehat{\varepsilon}^{\mathcal{N}}$, Figure \ref{fig:clauset_tpl_pl} illustrates the mean estimate of $\alpha$ for 5,000 samples drawn from the Pareto distribution, with $\alpha = 2.5$, $\varepsilon = 100$, and n = 2,500, plotted as a function of $\varepsilon$ for power law (purple) and truncated power law (green) distributions. 

To show differences in the methods for computing $\widehat{\varepsilon}^{\mathcal{N}}$, we performed the following simulation study. For each $N \in \{20,30,50,100,500\}$, we draw $N$ samples from the distribution
\begin{equation}
    \begin{aligned}
        f_\Lambda(\lambda) &\propto \begin{cases}
        \rho(\lambda ; \alpha, \varepsilon) & x \geq \varepsilon,\\
        \exp(-\alpha \lambda/\varepsilon) & x < \varepsilon,
        \end{cases}
    \end{aligned}
    \label{eq:simulated_data}
\end{equation}
with $\rho(\cdot)$ denoting the density function for a Pareto distribution, $\alpha = 2.5$ and $x_{\min} = 1$. Then we estimate $\varepsilon$ and $\alpha_{\text{MLE}}$ using the KS statistic, the AD statistic, and Kuiper's statistic.

\begin{figure}[htbp]
    \centering
    \includegraphics[scale=.29]{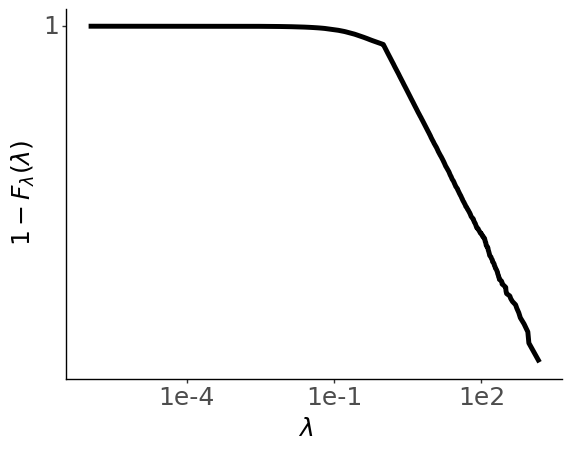}\includegraphics[scale=.29]{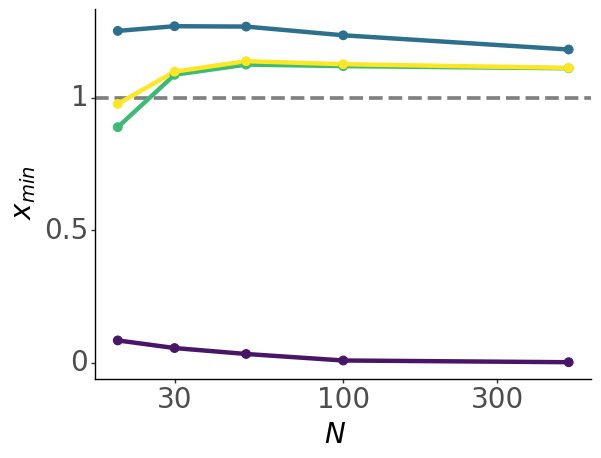}\includegraphics[scale=.29]{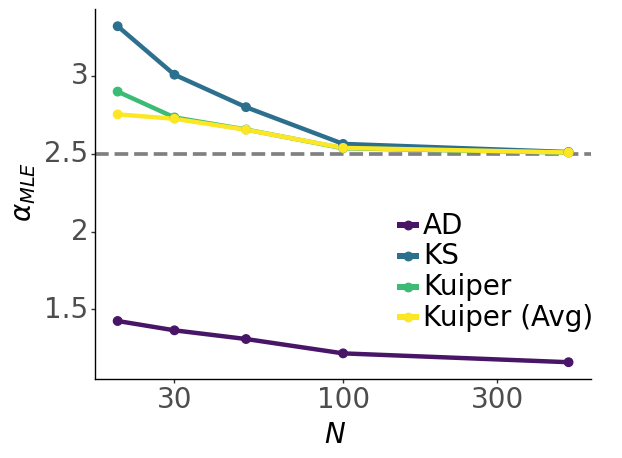}
    \caption{\textit{Left:} CCDF of 10,000 samples generated by Eq. \eqref{eq:simulated_data}. \textit{Middle:} Average fitted $\varepsilon$ by method and number of samples. \textit{Right:} average $\widehat{\alpha}_{\text{MLE}}$ by number of samples. 
    }
    \label{fig:example_simulated_data}
\end{figure}

The average values of $\varepsilon$ and $\alpha_{\text{MLE}}$ by sample size and metric are shown in Figure \ref{fig:example_simulated_data}. We find that the Kuiper statistic outperforms the KS statistic in estimation of power law parameters for all sample sizes, with more significant differences with smaller samples. Using neighborhood averaging with the Kuiper statistic does at least as well, producing better parameter estimates at the smallest sample sizes.

\begin{figure}[htbp]
    \centering
    \includegraphics[scale=.73]{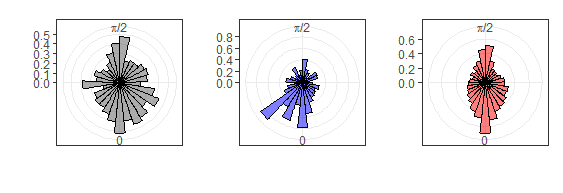}
    \caption{Radial histograms of jump angles for the individual devices in Figure \ref{fig:survival}  over the 12 week period.}
    \label{fig:jump_angles}
\end{figure}

\paragraph*{Pure Jump Angle Distribution}  A standard L\'evy flight assumes a uniform angular distribution of movement, but in practice the angular distribution of movement must also be specified. Figure \ref{fig:jump_angles} shows radial histograms of jump angles for three example individuals over the 12-week period, evidencing non-uniform traversal directions within individuals.

Due to its favorable conjugacy properties \citep{guttorp1988finding}, a convenient choice to model angular movement is the von Mises distribution with density
\begin{equation}
    \begin{aligned}
        f(\vartheta \mid \mu, \kappa) &= \{2\pi I_0(\kappa)\}^{-1} \exp\{\kappa \cos(\vartheta - \mu)\}, \qquad 0\leq \vartheta \leq 2\pi,\ \kappa \geq 0,\\
    \end{aligned}
    \label{eq:vmdist}
\end{equation}
where $\mu$ is the mean direction, $\kappa$ is a concentration parameter, and
\begin{equation*}
    \begin{aligned}
        I_0(\kappa) &= \pi^{-1} \int_0^{\pi} \exp(\kappa \cos(\phi))d\phi
    \end{aligned}
\end{equation*}
is the modified Bessel function of order zero \citep{abramowitz1965handbook}. When $\kappa = 0$ the von Mises distribution \eqref{eq:vmdist} becomes a uniform distribution in the unit circle and often serves as a reasonable and convenient prior. \citet{guttorp1988finding} showed that the conjugate prior of \eqref{eq:vmdist} is proportional to
\begin{equation*}
    \begin{aligned}
        f(\mu, \kappa \mid c, \mu_0, R_0) &\propto \{I_0(\kappa)\}^{-c}\exp\{\kappa R_0 \cos(\mu - \mu_0)\}.
    \end{aligned}
\end{equation*}
Given a set of observed angles $\underline{\vartheta} = (\vartheta_1,\ldots, \vartheta_n)$, the posterior distribution of $(\mu, \kappa)$ is
\begin{equation*}
    \begin{aligned}
        f(\mu, \kappa \mid \underline{\vartheta}) &\propto \{I_0(\kappa)\}^{-n-c}\exp\{\kappa R_n\cos(\mu - \mu_n)\},
    \end{aligned}
\end{equation*}
where 
\begin{equation*}
    \begin{aligned}
        R_n \cos \mu_n &= R_0\cos\mu_0 + \sum_i \cos\vartheta_i,\quad\,\mbox{and}\quad
        R_n \sin \mu_n &= R_0\sin\mu_0 + \sum_i \sin\vartheta_i.
    \end{aligned}
\end{equation*}
In what follows, we model the jump angle distribution $f_{\theta}(\theta ; \vartheta)$ as a von Mises distribution with common parameter $\vartheta$.

\subsection{Memory Dependence} Identifying clusters of repeated activity is an important aspect of understanding the long-term activity distribution of individuals. While the mixture model in Eq. \eqref{eq:lfcm} captures heterogeneous activities and distinguishes those activities from travel, it does not identify the regions in which the activities occur, nor it incorporates locations in the future likelihood of travel. In this section, we propose a probabilistic method for conditional identification of activity regions and returns to those regions.

Given a location $\mu_0$, the average Brownian movement over the time period $[0,T]$ is
\begin{equation}
    \begin{aligned}
     \tfrac{1}{T}\int_0^T \left[\underline{\mu}_0 + \underline{\mu} t + \sigma \underline{B}(t)\right]dt &\sim \mathcal{N}\left(\underline{\mu}_0 + \tfrac{T}{2}\text{diag}(\underline{\mu}), \tfrac{T}{3}\text{diag}(\sigma^2)\right).
    \end{aligned}
    \label{eq:bm_avg}
\end{equation}
We assume that the distribution of $\underline{x}(t_i)$ being a return to the activity region $z$ is proportional to $\mathcal{N}(\underline{\widetilde{\mu}}_z, \mathbf{\Sigma}_z)$, where $z$ represents a contiguous Brownian motion in the activity group $g$, and
\begin{eqnarray*}
\underline{\widetilde{\mu}}_z &\equiv& \underline{x}(t_{0,z}) + \tfrac{1}{2}T_z\text{diag}(\underline{\mu}_g),\\
\mathbf{\Sigma}_z &\equiv & \tfrac{1}{3}T_z\mathbf{\Sigma}_g^*,\\
T_z &\equiv &t_{n,z} - t_{0,z}.
\end{eqnarray*}

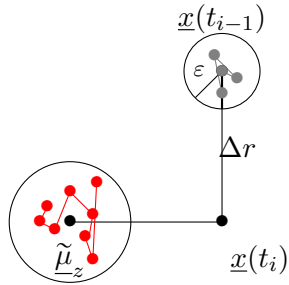
\begin{figure}[htbp]
\begin{center}
\begin{tikzpicture}
\draw(2,2.7) node{$\underline{x}(t_{i-1})$};
\draw(2.5,-.5) node{$\underline{x}(t_i)$};
\draw(0,-.5) node{$\underline{\widetilde{\mu}}_z$};
\draw(1.7,2) node{\footnotesize{$\varepsilon$}};
\draw(2.2,1) node{$\Delta r$};
\draw(2,2) circle (.5cm);
\draw(0,0) circle (.8cm);
\draw[-] (2,2) -- (2,0);
\draw[-] (2,2) -- (1.65,1.65);
\draw[-] (2,0) -- (0,0);
\draw[gray, -] (2,1.7) -- (1.9,2.2);
\draw[gray, -] (1.9,2.2) -- (2.2,1.9);
\draw[gray, -] (2.2,1.9) -- (2,2);
\draw[red, -] (-.3, .2) -- (-.4, 0);
\draw[red, -] (-.4, 0) -- (-.2, -.1);
\draw[red, -] (-.2, -.1) -- (0, .4);
\draw[red, -]  (0, .4)  -- (.3, .1);
\draw[red, -] (.3, .1) -- (.25, -.2);
\draw[red, -] (.2, -.2) -- (.3, -.5);
\draw[red, -] (.3, -.5) -- (.35, .52);

\draw[thick,-] (2,2) -- (2,1.5);
\node at (2,2) {\textbullet};
\node at (2,0) {\textbullet};
\node at (0,0) {\textbullet};
\node[gray] at (2, 1.7) {\textbullet};
\node[gray] at (1.9,2.2) {\textbullet};
\node[gray] at (2.2,1.9) {\textbullet};
\node[gray] at (2,2) {\textbullet};
\node[red] at (-.3, .2) {\textbullet};
\node[red] at (-.4, 0) {\textbullet};
\node[red] at (-.2, -.1) {\textbullet};
\node[red] at (0, .4) {\textbullet};
\node[red] at (.3, .1) {\textbullet};
\node[red] at (.2, -.2) {\textbullet};
\node[red] at (.3, -.5) {\textbullet};
\node[red] at (.35, .52) {\textbullet};
\end{tikzpicture}
\end{center}
\caption{An example hypothetical jump from $\underline{x}(t_{i-1})$ to $\underline{x}(t_i)$ relative to an activity region $z$ parameterized by ($\underline{\widetilde{\mu}}_z, \mathbf{\Sigma}_z$). An example sample path within the activity region is shown in red and.}
\label{fig:convol_ex}
\end{figure}

Figure \ref{fig:convol_ex} provides an illustrative example. Suppose that at some point in the past an individual was active in a region $z$ with center $\underline{\widetilde{\mu}}_z$ and covariance $\mathbf{\Sigma}_z$, denoted by the red sample path in the figure. At time $t_{i}$ an individual moves to $\underline{x}(t_i)$ according to the jump distribution $f_{J^*}(\Delta r ; \vartheta, \alpha)$. The conditional likelihood that the jump $\underline{x}(t_{i})$ is then classified as a return to activity $z$, i.e. $f_{z(t_i)}(z(t_i) | \underline{\Delta x}(t_i), \underline{\theta})$, is proportional to $\mathcal{N}(\widetilde{\underline{\mu}}_z,\mathbf{\Sigma}_z)$. Specifically,
\begin{equation*}
    \begin{aligned}
        f_{z(t_i)}(z(t_i)\mid \underline{\Delta x}(t_i),\underline{\theta}) &= \dfrac{\mathcal{N}(\widetilde{\underline{\mu}}_z,\mathbf{\Sigma}_z)}{\sum_{\zeta \in \mathcal{Z}_i}\mathcal{N}(\widetilde{\underline{\mu}}_\zeta,\mathbf{\Sigma}_\zeta)},
    \end{aligned}
\end{equation*}
where $\mathcal{Z}_i$ is the set of all activities identified at time $t_i$.

This serves as a computationally convenient conditional classification procedure for capturing nuances of individual mobility. Returns to activity regions must be defined explicitly since, although individuals do exhibit nonrandom routines, GPS spatial data measurements are on a continuous scale and unlikely to be repeated at multiple time points. When viewed in continuous space, the number of visited locations can resemble a L\'evy flight, but as a more intuitive notion of activity regions, the number of new locations visited decreases over time. This conditional identification procedure incorporates historical movement variability in activity areas during the estimation rather than relying on the assumption of uniform likelihoods within grid cells used in previous work \citep{chen2020measuring, alessandretti2017multi}. Once identified, these distributions, characterized by the parameters $(\widetilde{\underline{\mu}}_z, \mathbf{\Sigma}_z)$, can be used as known locations to simulate realistic probability-based return behavior observed in true mobility (e.g., using the 95\% CI of the distribution as a location of activity, see Figure \ref{fig:app123370}).

\subsection{The L\'evy Flight Cluster Model and its Bayesian specification} \label{sec:bayespostlfcm}

We are now ready to introduce our main modeling framework for human mobility patterns.

\begin{definition}[L\'evy Flight Cluster Model]
For a stochastic process $\{\underline{X}(t) : t\geq 0\}$, the \textit{L\'evy Flight Cluster Model (\texttt{LFCM})} is given by the mixture
\begin{equation}
    \begin{aligned}
            f_{\underline{\Delta x}}\left(\underline{\Delta X}(t) \mid N_G, \underline{\theta}, \Delta t \right) 
            &\equiv (1-\nu)\left[\sum_{g=1}^{N_G} \omega_g\mathcal{N}\left(\underline{\Delta x}(t) ; \underline{\mu}_g \Delta t, \mathbf{\Sigma_g^*}\Delta t \right)\right]\\
        &\phantom{\equiv }  +\nu f_{\Delta r}(\Delta r ; \alpha)f_{\Delta \theta}(\Delta \theta ;\vartheta)\\
        &\phantom{+ } \times\left[(1-\mathfrak{p}) + \mathfrak{p} \sum_{z=1}^{N_Z}\xi_z\mathcal{N}\left(\underline{\Delta x}(t); \underline{\widetilde{\mu}}_z, \mathbf{\Sigma}_z\right)\right],
    \end{aligned}
    \label{eq:main_model}
\end{equation}
where $\nu$ is the probability of a jump occurring, $N_G$ is the number of activity groups, $\omega_g$ is the probability that $\underline{\Delta x}(t)$ is part of activity $g$, $\mathfrak{p}$ is the probability of a return occurring, and $\xi_z$ is the conditional probability of returning to activity region $z$. Note that $N_Z$ is implicitly defined by the number of identified contiguous Brownian motion sample paths that belong to one of the $N_G$ activities.
\end{definition}

We define the $n \times 1$ latent vectors $\underline{b}, \underline{c}, \underline{\eta}$, and  $\underline{z}$: the assignment of observation $i$ to activity cluster $g$ is denoted by the latent vector $\underline{c} \in \{1,\ldots, N_G\}^n$; assignment to a jump step is denoted by the latent vector $\underline{b} \in \{0,1\}^n$, the assignment of a return step is denoted by the latent vector $\underline{\eta} \in \{0,1\}^n$, and the assignment to an activity area is denoted by $\underline{z} \in \{(g,l)\ :\ g=1,\ldots N_G;\ l = 1,\ldots, L_g\}$. The classification $\underline{z}$ represents a return to the activity space associated with the $l^{\text{th}}$ contiguous Brownian motion in activity $g$. Let $\underline{\theta} \equiv \{\underline{\mu}_g, \mathbf{\Sigma}_g^*, \alpha, \vartheta, \omega_g\}_{g=1}^G$ and $\underline{\mathfrak{h}} \equiv \{\underline{b}, \underline{c}, \underline{\eta}, \underline{z}\}$. The mixture likelihood in Eq. \eqref{eq:main_model} becomes
\begin{eqnarray}
    f_{\underline{\Delta x}, \underline{\mathfrak{h}}}\left(\underline{\Delta X}(t), \underline{\mathfrak{h}} \mid N_G, \underline{\theta}, \Delta t \right) 
            & \equiv & A^{\mathbf{1}\{b=0\}}B^{\mathbf{1}\{b=1\}},\label{eq:posterior}
\end{eqnarray}
\noindent where
\begin{eqnarray*}
    A & = & (1-\nu)\sum_{g=1}^{N_G}\left[ \omega_g\mathcal{N}\left(\underline{\Delta x}(t) ; \underline{\mu}_g \Delta t, \mathbf{\Sigma_g^*}\Delta t \right)\right]^{\mathbf{1}\{c=g\}},\quad \mbox{and}\\
    B & = & \nu f_{\Delta r}(\Delta r ; \alpha)f_{\Delta \theta}(\Delta \theta ;\vartheta) (1-\mathfrak{p})^{\mathbf{1}\{\eta=0\}}
        \left[\mathfrak{p} \sum_{z=1}^{N_Z}\left[\xi_z\mathcal{N}\left(\underline{\Delta x}(t); \underline{\widetilde{\mu}}_z, \mathbf{\Sigma}_z\right)\right]^{\mathbf{1}\{z=(g,l)\}}\right]^{\mathbf{1}\{\eta = 1\}}.
\end{eqnarray*}

A Bayesian framework for inference in the \texttt{LFCM} is defined by the prior specification given in Table \ref{table:model_params} and follows in part the priors for finite latent mixtures parameters of \citet{ryan2017bayesian}. These priors are chosen to be conjugate and non-informative to lead to efficient MCMC samplers from the corresponding joint posterior distributions. Nuisance parameters can be integrated out as needed. \citet{ryan2017bayesian} find the prior distribution $N_G \sim \text{Gamma}(\tfrac{1}{2},\tfrac{1}{2})$ works well in practice for absorbing existing components and ejecting new components in a finite mixture. The minimal jump distance $\varepsilon$ for classification as a jump governs the scale of the model, and is assumed to be known to obtain conjugacy. The determination of appropriate values for $\varepsilon$ is discussed in Sections \ref{sec:sim} and \ref{sec:appl}. 

\begin{table}[htbp]
\begin{center}
\begin{tabular}{*4l}
\toprule
   \textbf{Parameter} &  \textbf{Prior Distribution} &  \textbf{Hyper-prior Distribution} \\ 
\midrule
    $N_G$ & Poiss($\lambda$) & Gamma($\frac{1}{2}, \frac{1}{2}$) \\
    c & Multin($\underline{\omega}$) & $\underline{\omega} \sim$ Dir($\underline{1}$) \\ 
    $b$ & Bern($\nu$)  & $\nu \sim $ Beta$(2, 2)$ \\ 
    $\|\underline{\Delta x}\|_2\ |\ b = 0$ & $\mathcal{N}(\underline{\mu}_g \Delta t, \mathbf{\Omega}_g^{-1}\Delta t$) & $\underline{\mu}_g, \mathbf{\Omega}_g \sim NW(\underline{0}, .01, \frac{1}{2}, \frac{1}{2}\mathbf{I})$ \\ 
    $z$ & Multin($\underline{\xi}$)  & $\underline{\xi} \sim \text{Dir}(\underline{1})$ \\ 
    $\eta$ & Bern($\mathpzc{p}$) & $\mathpzc{p} \sim $ Beta($2, 2$) \\ 
    $\|\underline{\Delta x}\|_2\ |\ b = 1$ & Par($\alpha_g$) ($\varepsilon$ known)  &  $\alpha_g \sim $ Gamma$(\tfrac{1}{2}, \tfrac{1}{2})$ \\ 
    $\tan^{-1}\left(\frac{\Delta x_2}{\Delta x_1}\right)$ & VM($\vartheta$, $\tau$) & $\vartheta \sim $ VMises(0, 1) \\ 
    \hline
    \end{tabular}
    \caption{Prior specification for the \texttt{LFCM}.}
    \label{table:model_params}
    \end{center}
\end{table}

When interest lies only in activity regions, a collapsed MCMC sampler can be derived by integrating almost all parameters of the \texttt{LFCM} \--- see Appendix \ref{appendix:collapsedmcmc}. 

\section{Simulation study} \label{sec:sim}
We evaluate the efficacy of the \texttt{LFCM} model to represent the distribution of activity of an individual by simulating the trajectory of a simple routine in which, on a daily basis, an individual goes from home, to work, to a public space, and back home. With varying amounts of observed data, we show estimates of activity distributions for this example individual using the \texttt{LFCM} and a competing estimator called the conservative proportional time (CPT) estimator that is constructed conditional on a spatial grid \citep{dong2020statistical}. We show that the \texttt{LFCM} consistently captures the three key locations visited by the example individual, while the estimates produced by the CPT estimator exhibit more variability because they depend on the size of the cells in the spatial grid used. In addition, we show that the Pareto hyperparameter associated with minimal jump size does effect the behavior of the \texttt{LFCM}---in particular, it may incorrectly identify a jump process as an activity space when set to large values.

\paragraph*{Data Generation} We simulate the spatiotemporal trajectory of an individual during their activities of daily living. They first spend their time at a ``home'' location from 0:00 to 8:40 represented by the coordinates $(0,0)$. They then travel for 20 minutes to a ``work'' location at the coordinates $(1,1)$, and stay there from 9:00 to 17:00. They then they travel for 15 minutes to a public space at coordinates $(1,0)$, and stay there from 17:15 to 23:45. Finally, they take 15 minutes to return home. When at home and at work, the individual moves around the area according to a zero-drift Brownian motion $\sigma B(t)$ with standard deviation $\sigma = 0.1$. When in public space, the individual moves similarly with $\sigma=0.2$. The travel between these three key locations occurs on the shortest line between them. 

\begin{figure}[htbp]
    \centering
    \includegraphics[scale=.9]{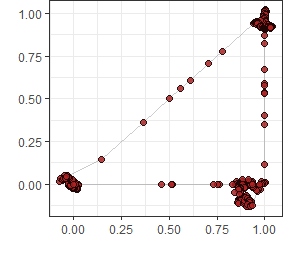}
    \caption{Simulated GPS location data.}
    \label{fig:simulation}
\end{figure}

We generate a new GPS location for the example individual every 10 minutes when at home and at work, every 5 minutes when in public space, and every 60 seconds when traveling between any two key locations. This reflects a real-world location data collection process in which individuals are more likely to use GPS tracking applications during high-movement activities (e.g., fitness tracking) or travel compared to time periods with low or no spatial movement. GPS locations during travel are simulated from a uniform distribution between the source and destination to simulate the possibility of traffic. Figure \ref{fig:simulation} displays the generated data and illustrates the simulated routine. 
We generate three subsets of the simulated data shown in Figure \ref{fig:simulation}: the entire set of simulated data, a random subset consisting of 50\% of the simulated data, and a random subset consisting of 25\% of the simulated data. We used these three subsets to estimate the activity distributions with LFCM together with the CPT estimator of \citet{dong2020statistical}.

\paragraph*{The Conservative Proportional Time Estimator (CPT)} The CPT estimator is a discrete estimator of an individual's activity distribution. We assume that a spatial window $\mathcal{A}$ in which the individual is active is divided into cells $\mathcal{C}$ of equal size. Denote by $\mathcal{X}=\{\underline{x}(t_j): j=1,\ldots,n\}$ the GPS locations of this individual. The probability that the individual is located in the grid cell $a\in \mathcal{C}$ is proportional to the time spent in this grid cell \citep{dong2020statistical}:
\begin{equation}
    \begin{aligned}
        \pi_a &= \dfrac{\sum\limits_{j=2}^n \Delta t_{j} \mathbf{1}_{\{\underline{x}(t_j)\in a,\  \underline{x}(t_{j-1}) \in a\}}}{\sum\limits_{a^{\prime}\in \mathcal{C}}\sum\limits_{j=2}^n \Delta t_{j}\mathbf{1}_{\{\underline{x}(t_j)\in a^{\prime},\  \underline{x}(t_{j-1}) \in a^{\prime}\}}},
    \end{aligned}
    \label{eq:c2cpt}
\end{equation}
This estimator is conservative because the time spent traveling between two grid cells is excluded from the calculation. \citet{chen2020measuring} define the density ranking $\widehat{\alpha}(\underline{x})$ as
\begin{equation}
    \begin{aligned}
        \widehat{\alpha}(\underline{x}) &= \frac{1}{n} \sum_{i=1}^n \mathbf{1}\{\widehat{p}(\underline{X}_i) \leq \widehat{p}(\underline{x})\},
    \end{aligned}
    \label{eq:density_ranking}
\end{equation}
where $\widehat{p}(\underline{x})$ is defined as $\Sigma_{a\in\mathcal{C}} \pi_a \mathbf{1}\{\underline{x}\in a\}$. The density ranking function $\widehat{\alpha}(\underline{x})$ is the fraction of observations in $\mathcal{X}$ whose estimated density is lower than the estimated density of the given point $\underline{x}$. The density ranking of a point $\underline{x}$ is interpreted as $\underline{x}$ being in the top $100\times [1 - \widehat{\alpha}(\underline{x})]\%$ of locations that an individual is observed visiting. This can be used to generate level sets of the form
$$\widehat{A}_\gamma = \{\underline{x}:\widehat{\alpha}(\underline{x}) \geq 1 - \gamma\},$$
corresponding to the top $\gamma\times 100\%$ activity regions.

\begin{figure}[htbp]
    \includegraphics[scale=.5]{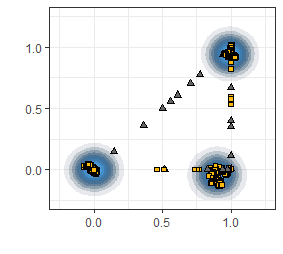}\includegraphics[scale=.5]{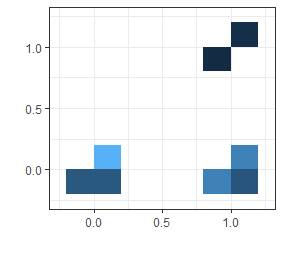}\\
    \includegraphics[scale=.5]{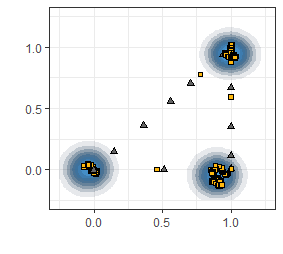}\includegraphics[scale=.5]{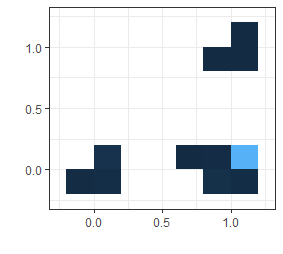}\\
    \includegraphics[scale=.5]{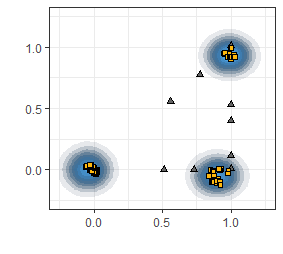}\includegraphics[scale=.5]{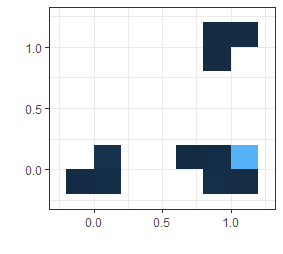}\\
    \caption{Activity space estimates LFCM classification of simulated data (left) and the kernel density estimator \citep{chen2020measuring} is displayed as a reference (right). The subsets displayed are 100\% (top), 50\% (middle), and 25\% (bottom) of the original data.}
    \label{fig:sim_setup}
\end{figure}

\paragraph*{Results} We used the MCMC Algorithm \ref{algo:gibbs} 
described in Appendix \ref{appendix:collapsedmcmc} to estimate activity types, jump points, and activity regions with the \texttt{LFCM}. As a point of comparison, the level set $\widehat{A}_{0.9}$ is estimated using the CPT in Eq. \eqref{eq:c2cpt}. Figure \ref{fig:sim_setup} displays the results of LFCM (left) where Brownian motion is identified by square points, jumps are identified by triangular points, and shaded areas are identified activity regions; and the CPT method of \citet{dong2020statistical} (right). The top row corresponds to all data, the middle row corresponds to a random selection of 50\% of the data, and the bottom row corresponds to a random selection of 25\% of the data.

\begin{figure}
    \centering
    \hspace{-.3in}\includegraphics[scale=.5]{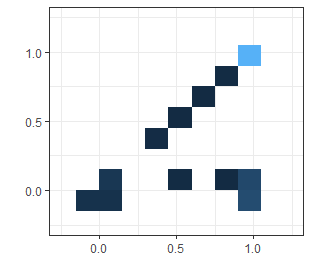}\includegraphics[scale=.55]{cpt_100.png}\includegraphics[scale=.5]{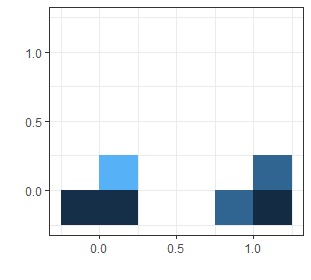}
    \caption{Proportional Time estimates with grid size 0.15 (left), 0.20 (middle), and 0.25 (right).}
    \label{fig:grid_size}
\end{figure}

While the \texttt{LFCM} and the CPT recover the activity regions associated with home $(0,0)$, work $(1,1)$ and public space $(1,0)$, \texttt{LFCM} achieves a Jensen-Shannon divergence of $1.06\times 10^{-5}$ relative to the generative distribution, while CPT has a Jensen-Shannon divergence of 0.013. Figure \ref{fig:sim_setup} shows the sensitivity of CPT  \citep{dong2020statistical} to the implicit grid size. Indeed, changes in grid size can have substantial effects on the activity regions identified. Figure \ref{fig:grid_size} illustrates three distinct grid sizes that substantively change the level set $\widehat{A}_{0.9}$. From left to right, the grid sizes used are 0.15, 0.20, and 0.25. A grid size of 0.15 identifies additional jump points as activity regions while shrinking the area of true activity, while a grid size of 0.25 removes one of the three areas of activity altogether.

\begin{figure}[htbp]
    \centering
    \includegraphics[scale=.52]{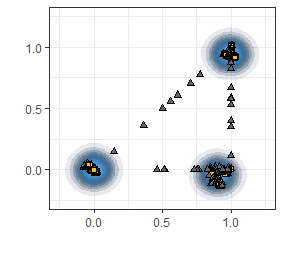}\includegraphics[scale=.52]{simulation_results.png}\includegraphics[scale=0.47]{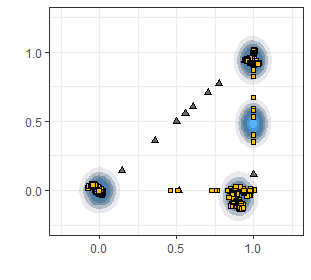}
    \caption{\texttt{LFCM} with minimal jump distance parameter $\varepsilon=0.01$ (left panel), $0.10$ (middle panel), and $0.20$ (right panel).}
    \label{fig:epsilon_diffs}
\end{figure}

\texttt{LFCM} is subject to less sensitivity, as the majority of the parameters in the \texttt{LFCM} can be analytically integrated out. However, \texttt{LFCM} has sensitivity to the \textit{a priori} minimal jump length parameter $\varepsilon$. The simulation above uses a minimal jump length value $\varepsilon$ of $0.1$ based upon visual inspection of the data. If this value is set low, a number of points in the activity space may be labeled as jumps, while if the value is high, origin-destination travel may be labeled as Brownian motion activity. Figure \ref{fig:epsilon_diffs} illustrates this behavior. The left panel shows \texttt{LFCM} results on the simulated data when $\varepsilon=0.01$, the middle panel shows \texttt{LFCM} results with $\varepsilon=0.10$, and the right panel shows \texttt{LFCM} results with $\varepsilon=0.20$. In the left panel, a substantial number of the Brownian motion points at $(1,0)$ are instead classified as jump points, while in the right panel, a portion of travel is classified as an activity space. For this reason, a careful calibration of the minimal jump length is needed by fitting the \texttt{LFCM} on a grid of possible values for $\varepsilon$.

\begin{figure}[htbp]
    \centering
    \includegraphics[scale=.8]{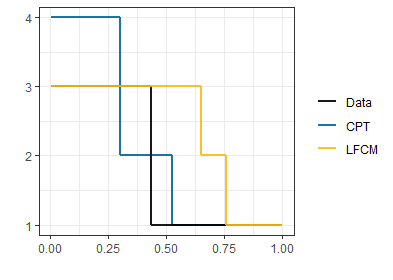}
    \caption{Persistence curve of activity regions for the data (black), CPT (yellow), and LFCM (red).}
    \label{fig:persistence_curve}
\end{figure}

\paragraph*{Persistence Curves} The activity regions estimated by the \texttt{LFCM} naturally provide a probability density from the mixture of Brownian motions which generate them. We use this estimated density, along with the discrete density estimated with the CPT estimator to generate the \textit{persistence curve} in Figure \ref{fig:persistence_curve}, which illustrates the level sets $\widehat{A}_\gamma$ for $\gamma \in [0,1]$.
 The persistence of a connected component is defined by its birth and death times, which are determined by the corresponding level of density ranking. A decrease in the number of connected components suggests that either a connected component disappeared or two components merged. The persistence curve also provides evidence of mode stability; modes that persist over a larger range of $\gamma$ values are more robust, suggesting that they represent a true feature of the underlying data distribution. While the CPT identifies 4 components, the LFCM identifies three individual connected components similarly to the original data generation mechanism. Moreover, each of the modes identified by the CPT has a lower persistence than those identified by the \texttt{LFCM}, suggesting the latter provides a more stable estimate of the true activity spaces.

\section{Analysis of mobile device GPS data} \label{sec:appl}

In this section we provide an application of the \texttt{LFCM} to real GPS data of the individual devices described in Section \ref{sec:gps}. First, to qualitatively illustrate the estimation output, we use the \texttt{LFCM} to show the types of activity and the activity activity spaces for a single device. Second, to demonstrate the \texttt{LFCM}'s efficacy of anonymized path generation, for the same device we display the generated sample path using the identified activity regions and draws of the posterior parameters returned by the model. Third, we provide a quantitative comparison across all weeks and individual devices between the generated sample paths and the true mobility data on the mobility metrics reviewed in Section \ref{sec:metrics}: jump length, mean squared displacement (MSD), radius of gyration and frequented locations. Fourth, we show that the proportion of new locations visited over time of the simulated trajectories are aligned with prior findings in the literature. Fifth, we qualitatively and quantitatively assess the stability of activity spaces identified by the \texttt{LFCM} by estimating the model over progressively longer time intervals. Finally, conditional on the known trajectories, we use the \texttt{LFCM} estimate to simulate and interpolate a continuous trajectory for each individual, allowing for direct comparison across non-uniformly recorded trajectories, showing that the \texttt{LFCM} can be used to estimate activity overlap.

\begin{figure}[htbp]
        \centering
            \includegraphics[scale=.7]{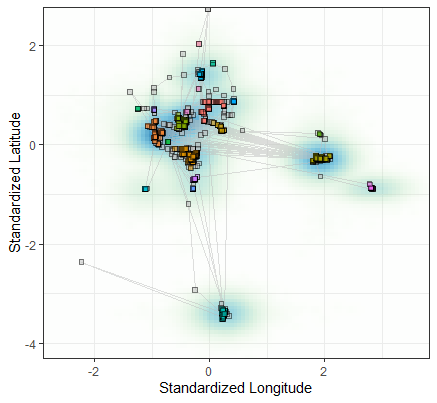}
        \caption{Example of \texttt{LFCM} on 4 weeks of human mobility data. Colored points represent distinct activities, gray points represent jumps, and shaded areas represent activity regions, according to the LFCM estimate.}
        \label{fig:app123370}
\end{figure}

\paragraph*{Activity region estimation} To obtain draws of posterior parameters for an example individual device's activity space, we obtain posterior samples of the parameters of the \texttt{LFCM} using 10,000 iterations of Algorithm \ref{algo:gibbs} described in Appendix \ref{appendix:collapsedmcmc} based on their 1,116 GPS locations recorded in four weeks. We generate activity regions using uncertainty bounds for the individual's unobserved time given the posterior Brownian motion parameters $(\underline{\mu}_g, \mathbf{\Sigma}_g^*)$ of an activity. 
Figure \ref{fig:app123370} displays the entire movement of the example individual device over the four-week period. The colored points represent contiguous Brownian paths within the same activity group $g$, and the shaded regions represent the activity regions.
 Each of the gray points in Figure \ref{fig:app123370} represents a location record identified as a non-Brownian jump in the observed process, i.e. as having a low probability of being a member of any of the activity groups after accounting for the time between records, usually occurring when a point is spread spatially far from neighboring data. The number of jumps is controlled by the hyperparameters $(\alpha, \varepsilon)$ of the Pareto distribution. 
 The parameter $\varepsilon$ controls the minimum distance from the prior location that can be identified as a jump, and the parameter $\alpha$ controls the probability concentration that a far point is classified as a jump relative to a Brownian motion.

The shaded regions in Figure \ref{fig:app123370} are activity regions using LFCM representing the potential range of the device's movement between records while in the area after accounting for drift and dispersion. Specifically, the region represents a 95\% confidence region assuming the Brownian bridge movement model in Eq. \eqref{eq:bbmm} with dispersion $\mathbf{\Sigma}_g^*$ and time $\Delta t$ between each point identified as in the same activity regions. Activity regions may appear large for two reasons: the magnitude of the dispersion parameter within the activity group $g$ is high, or the time spent in the area is high. The constructed activity regions need not be elliptical, but activities with low dispersion over a long period of time tend to create elliptical regions of uncertainty.

\paragraph*{Model record generation} We show that simulations from the \texttt{LFCM} can be used to generate representative trajectories with characteristics that match the mobility of the individual device. Based on posterior samples from the \texttt{LFCM}, we estimate the jump length distribution, the activity type distribution, the mean and variance of a Brownian motion describing the activities' movement. Areas of frequent return are a function of these quantities. 

\begin{figure}[htbp]
    \centering
    \includegraphics[scale=.7]{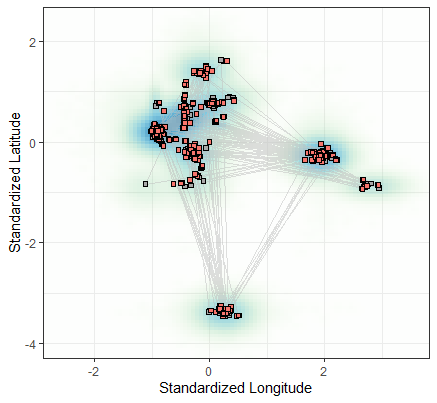}
    \caption{Simulated movement for an individual device based on the LFCM MAP estimate.}
    \label{fig:four_sims}
\end{figure}

Using one week of data from the individual device described above, we draw samples from the joint posterior distribution of the \texttt{LFCM}. We use these samples to determine MAP parameter estimates, identify activity regions and generate sample paths. Figure \ref{fig:four_sims} displays the trajectory of the device in 500 points simulated using the MAP estimates, where Brownian motion movement is shown in red, L\'{e}vy jumps are shown in gray, and the identified activity regions are shaded in blue. This simulation can be compared with Figure \ref{fig:app123370}, which is the true mobility of the device in the same week---the simulated points appear to be very similar to those in Figure \ref{fig:app123370}, which are the true mobility data.

This simulation process also results in reasonable mobility metrics mentioned in Section \ref{sec:metrics}. Table \ref{tab:metrics} displays the mean 
 and standard deviation of jump length, MSD, and radius of gyration across 12 weeks of movement and 293 individual devices for the observed data, data simulated from \texttt{LFCM}, and data simulated from the conservative proportional time estimator in Eq. \eqref{eq:c2cpt}. Whereas the \texttt{LFCM} closely reproduces jump length, MSD, and radius of gyration, simulations from the CPT estimator tend to exhibit under-bias on mean jump length and MSD.

The noticably better performance of the LFCM compared to the CPT estimator can be explained by a couple of factors. First, the likelihood of movement is constant within grid cell, and some cells are much more likely than others. This causes a large amount of the trajectory to be within-cell movement. Second, region size is uncorrelated with time spent in a region for the CPT estimator. In contrast, for the LFCM, with the exception of certain key locations, the likelihood of a region is correlated with the size of the region; nearby areas of return tend to merge if there is sufficient Brownian motion activity in those areas.
\begin{table}[htbp]
    \centering
    \begin{tabular}{lcccc}
        \toprule
              & Jump Length & MSD & Radius of Gyration \\
              \cmidrule{2-5}
         Observed &  1.556 & 2.889 & 1.250 \\
          & (0.052) & (0.041) & (0.0040) & \\
         \cmidrule(lr){2-4}
         Simulated (LFCM) & 1.580  & 2.975 & 1.190 \\
         & (0.035) & (0.056)  & (0.038) \\
         \cmidrule(lr){2-4}
         Simulated (CPT), W=0.2 & 1.120 & 1.136 & 0.902 \\
                                & (0.045) & (0.052) & (0.034) \\
         Simulated (CPT), W=1.0 & 1.305 & 1.187 & 0.935 \\
                                & (0.073) & (0.096) & (0.038) \\
         Simulated (CPT), W=1.5 & 1.599 & 1.589 & 1.140 \\
                                & (0.040) & (0.093) & (0.027)\\
         \bottomrule
    \end{tabular}
    \caption{Average jump length, MSD, and ROG for observed data, simulated data using LFCM, and simulated data using the CPT estimator. Standard deviations are in parentheses.}
    \label{tab:metrics}
\end{table}

\begin{figure}[htbp]
    \centering
    \includegraphics[scale=.8]{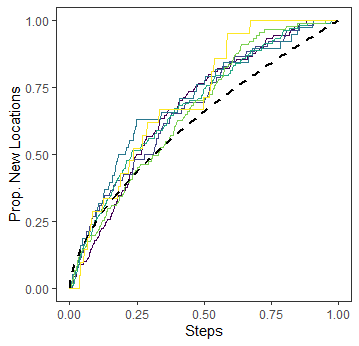}
    \caption{Representative figure for the LFCM simulated proportion of new locations visited over time for six simulated devices (colored) relative to $f(t) = t^{0.6}$ (dashed), the proportion of new locations visited estimated in \citep{song2010modelling}.}
    \label{fig:new_locs}
\end{figure}

\paragraph*{Frequented Locations} One of the primary difficulties in the estimation of individual mobility using L\'evy flight models is faithful replication of an individual's frequented locations. This is often visually assessed in the literature as the number of new locations an individual visits over the time period of data collection. Similarly, we plot the proportion of new locations visited over a week-long period of data collection--temporally normalized to the unit interval $[0,1]$--estimated by the \texttt{LFCM} in Figure \ref{fig:new_locs}; using MAP estimates estimated by the LFCM for six individual devices, we generate new trajectories conditioning on the observed times seen in the data. The figure shows a reference line of $t^{0.6}$, corresponding to the observed rates of exploration in the literature \citep{song2010modelling}, while the average across our simulations is $t^{0.52}$.

\begin{figure}[htbp]
    \centering
    \includegraphics[scale=.35]{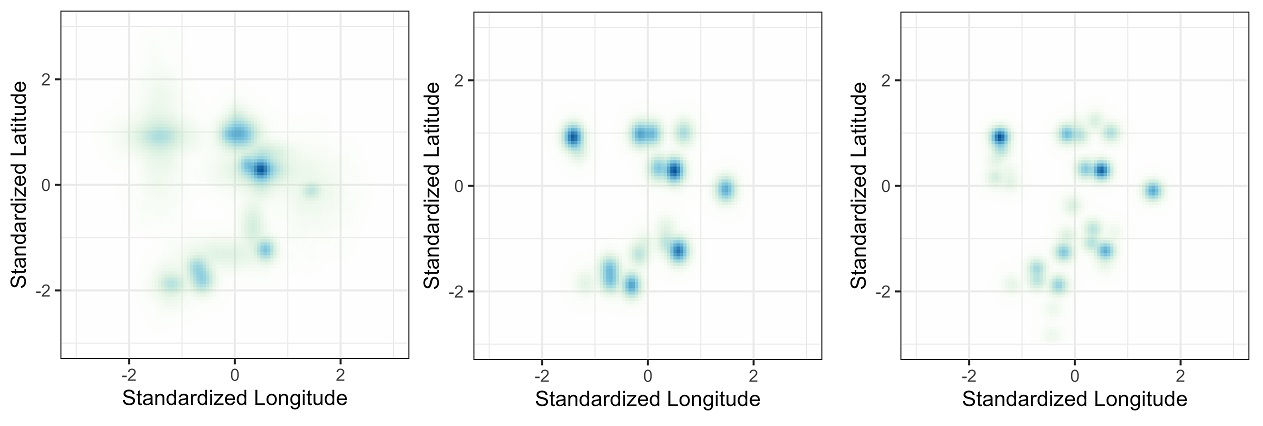}
    \caption{LFCM-based activity space distributions for an individual device after 1 week (left), 1 month (middle), and 3 months (right).}
    \label{act_over_time}
\end{figure}

\paragraph*{Activity Space Stability} While individuals might continually visit new locations, the number of places they spend most of their time stays relatively stagnant. We illustrate this in Figure \ref{act_over_time}, which displays an individual device's activity spaces estimated by \texttt{LFCM} after 1 week (left), 4 weeks (middle), and 12 weeks (right) of observation. While the individual's activity distribution becomes more concentrated, and the areas in which the individual spends the majority of their time becomes more obvious over time, the most frequented activity spaces appear to be stable after a single week; the activity regions around $(-1.5,1)$, $(0,0)$, and the multiple regions between $(-0.5,-2)$ and $(0.5,-1)$ are present across all estimates.

Figure \ref{act_over_time} also shows that along with persistent areas, new areas are identified over time. However, adults living in urban areas tend to have two or three most relevant places: work, home, and a primary social activity \citep{oldenburg1982third,finlay2019closure}. Although work may require significant travel, the few areas of highest activity are likely to be the most relevant to understanding patterns of habitual activity. To assess the extent \texttt{LFCM} can capture this behavior, we empirically evaluate convergence of the ``Top 1'' and ``Top 3'' activity areas as a function of the observed time. 

\begin{figure}[htbp]
    \centering
    \includegraphics[scale=.33]{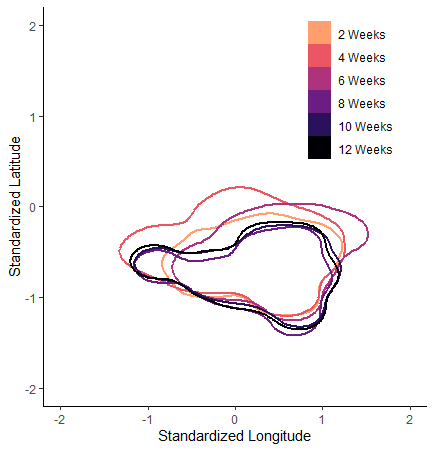}
    \includegraphics[scale=.33]{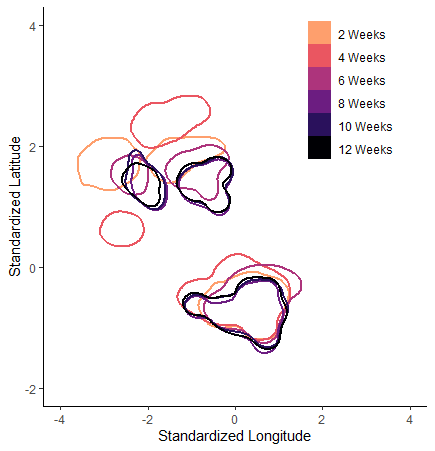}
    \includegraphics[scale=.33]{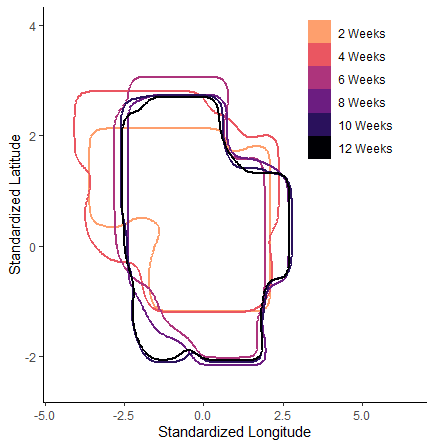}
    \caption{``Top 1'' activity area (left),``Top 3'' activity areas (middle), and the 80\% activity region (right), estimated using the LFCM with $w \in \{2,4,\ldots,12\}$ weeks of data.}
    \label{fig:activity_clusters}
\end{figure}

Specifically, let $\widehat{p}_{i,w}$ denote the estimated activity distribution for device $i$, estimated on points $\underline{x}_1,\ldots,\underline{x}_w$. First we define the $\gamma$ upper level set of distribution $\widehat{p}_{i,w}$ as
\begin{equation}
    \begin{aligned}
        \widehat{\mathcal{A}}_{i,w}^{\gamma} &\equiv \text{inf}\left\{ A \subset \mathbb{R}^2\ :\ \int_A \widehat{p}_{i,w}(\underline{x})d\underline{x} \geq 1 - {\gamma} \right\}.
    \end{aligned}
    \label{eq:activity_level_set}
\end{equation}
For each device we use the \texttt{LFCM} to estimate $\widehat{\mathcal{A}}_{i,w}^{0.8}$ in Eq. \eqref{eq:activity_level_set} from $w \in \{2,4,\ldots,12\}$ weeks of data points. For each device and week, we select the ``Top 1'' and ``Top 3'' disconnected subsets $A \in \widehat{\mathcal{A}}_{i,w}^{0.8}$, according to highest probability mass. Figure \ref{fig:activity_clusters} displays activity areas estimated with increasingly longer temporal data for an individual device. Both the ``Top 1'' and ``Top 3'' activity area estimates anecdotally corroborate the patterns shown in Figure \ref{act_over_time}; while some spatial error appears in the ``Top 3'' activity areas at week 4, the activity areas are generally consistent with even two weeks of data.

Finally, to construct a quantitative measure of convergence for the \texttt{LFCM} activity space estimates, we use the \texttt{LFCM} to estimate $\widehat{\mathcal{A}}_{i,12}^{0.8}$ on individual devices' full 12 weeks of data, and their corresponding ``Top 1'' and ``Top 3'' activity areas, as the ``true'' activity space. Then, for each device, we estimate the similarity between identified activity areas at week 12 and at week $w \in \{2,4,6,8,10\}$. Several similarity metrics described in Appendix \ref{appendix:similaritymetrics} 
can be applied to assess convergence in the estimated activity areas: the KL divergence, the Hausdorff distance, the Jaccard distance, and a modification of the Jaccard distance which we call the Overlap. The choice of metric depends on the representation and modeling of these areas. Generally, the Overlap and Jaccard distances are likely to be the least sensitive, especially to minor alterations in large activity spaces.

\begin{figure}[htbp]
    \centering
    \includegraphics[scale=.3]{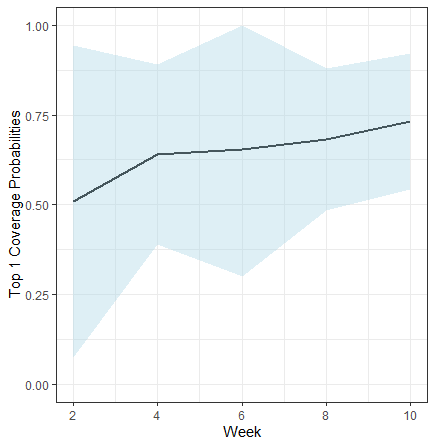}
    \includegraphics[scale=.3]{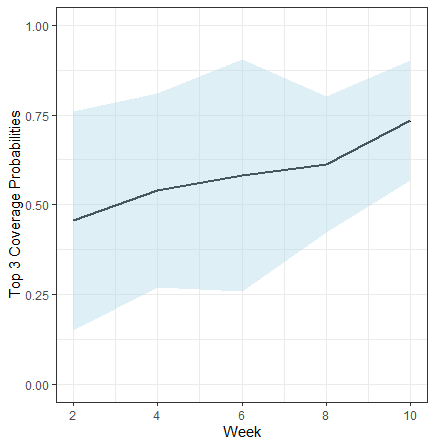}
    \includegraphics[scale=.3]{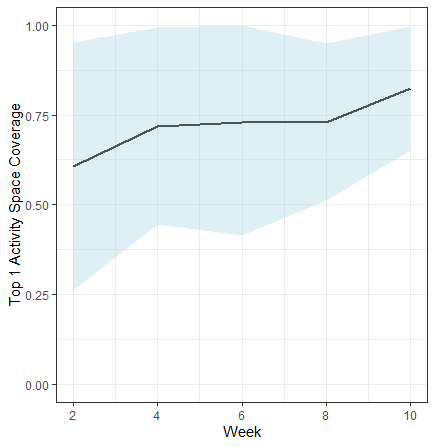}
    \caption{Mean proportion overlap averaged across devices for ``Top 1'' activity area (left), ``Top 3'' activity areas (middle), and 80\% activity area (right).}
    \label{fig:activity_overlap}
\end{figure}

The empirical mean and 95\% confidence band of the Overlap distance across devices for each week $w$ are shown in Figure \ref{fig:activity_overlap}. We found that, on average, ``Top 1'' and ``Top 3'' activity areas estimated on 2 weeks of data have approximately 50\% overlap with the corresponding estimates on 12 weeks of data. Using 4 weeks of data yields an average 10\% increase in overlap for both the ``Top 1'' and ``Top 3'' activity areas --- showing approximately 60\% overlap with 12-week activity areas. Finally, plots for both ``Top 1'' and ``Top 3'' show confidence interval lower bounds of approximately 20\% for 2-week estimates, suggesting substantial overlap with top activity areas estimated on the full 12 week data even in the worst case.

\paragraph*{Probabilistic Contact Networks} In social and behavioral sciences, understanding the probabilistic contact networks can aid in studying social interaction patterns, shared behaviors, and commonalities in lifestyle choices; in public health, these networks may imply similar exposures to health risks (e.g., pollution, noise), which can be important when studying disease spread or common health issues in a community; in urban planning and development, understanding these networks can inform the planning of public amenities, recreational areas, and even residential or commercial zones based on the population's behaviors and patterns of movement. In what follows, we leverage Bayesian parameter estimates of the \texttt{LFCM} to calculate these probabilistic contact networks.

\begin{figure}
    \centering
    \includegraphics[width=0.7\linewidth]{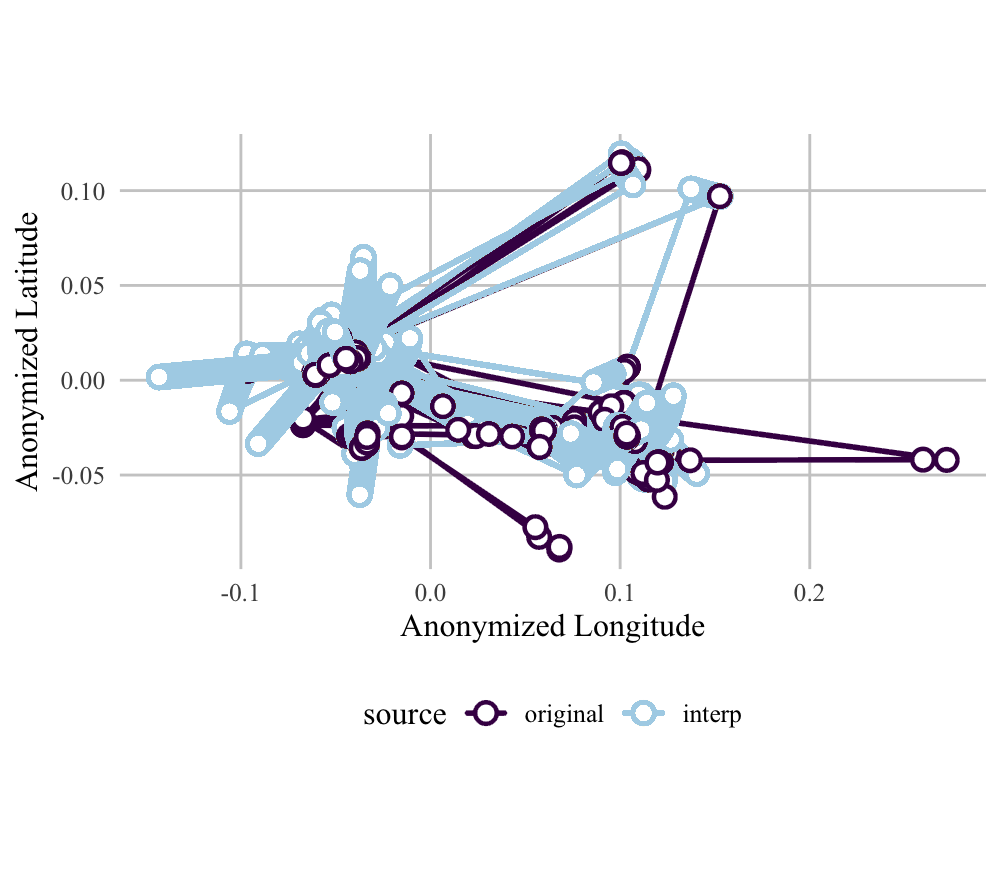}
    \caption{Example interpolated and extrapolated trajectory for individual device trajectory across one week. Latitude and longitude have been anonymized.}
    \label{fig:anon-interp}
\end{figure}

First, across all individual devices $i$ and weeks $w$, we find the minimum and maximum times across individuals for that week.
\[
t_{i,w}^{\mathrm{start}} = \min \{ t \in \text{week } w \mid \text{individual } i \text{ is observed at time } t \},
\]
and the final observation time as
\[
T_{i,w}^{\mathrm{end}} = \max \{ t \in \text{Week } w \mid \text{individual } i \text{ is observed at time } t \}.
\]
Then, the minimal initial observation time across all individuals in week \(w\) is given by
\[
t_{w}^{\mathrm{min}} = \min_{i} t_{i,w}^{\mathrm{start}},
\]
and the maximal end observation time is
\[
T_{w}^{\mathrm{max}} = \max_{i} T_{i,w}^{\mathrm{end}}.
\]
For each individual, interpolation and extrapolation are done by generating conditional sample paths from posterior scans of the $\texttt{LFCM}$---sampling randomly from the sequence of scans
\[
\theta^{(n_1)}, \theta^{(n_2)}, \ldots, \theta^{(N)}
\]
from the target posterior distribution \( p(\theta \mid D) \), after convergence is achieved. Figure \ref{fig:anon-interp} illustrates the resulting trajectory for one individual and week.

As a baseline comparison to this method, we also generate a simple interpolation and extrapolation strategy, which linearly interpolates between observed records, and extended to $t_{w}^{\min}$ and $T_{w}^{\max}$ by assuming that the individual remains still in their first and last observed record, respectively. 

For each scan, we then calculate the $L_2$ distance between individual trajectories across all devices for a given week, resulting in an activity overlap matrix. We generate 500 such matrices using 500 randomly sampled posterior scans of the individuals, which corresponds to a potential trajectory that the individual might follow over the observed period. 

\begin{figure}[htbp]
    \centering
    \includegraphics[width=0.45\linewidth]{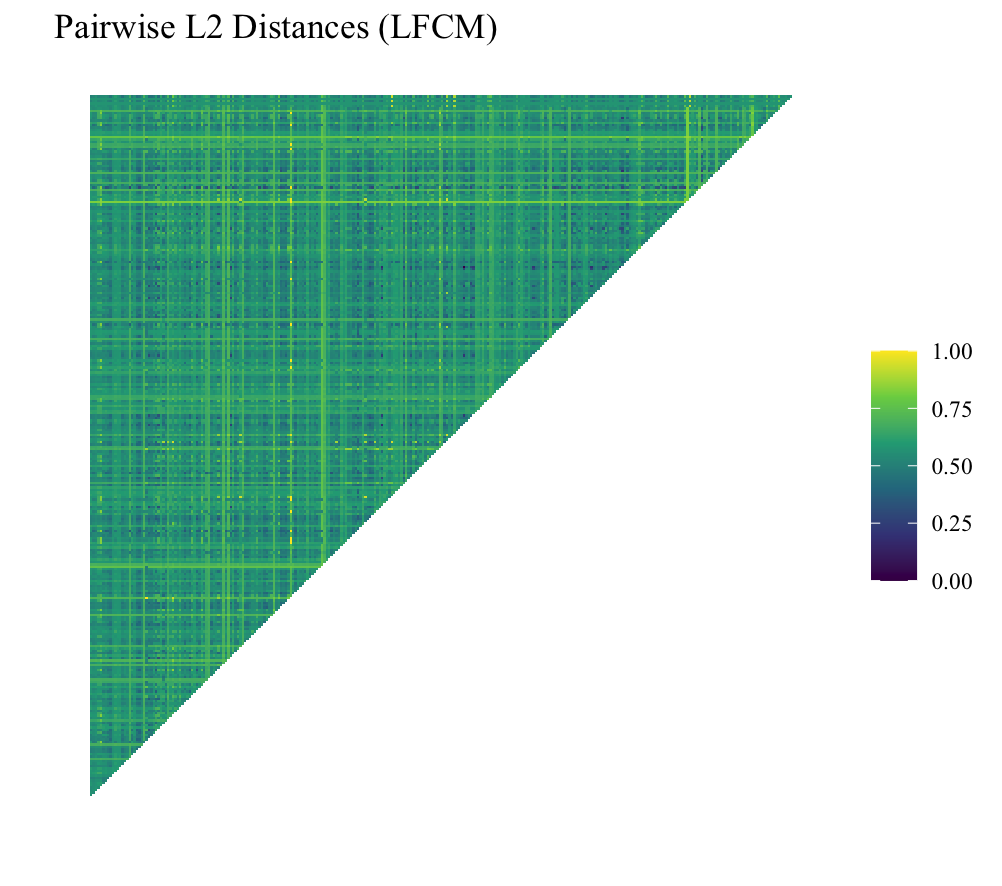}
    \includegraphics[width=0.45\linewidth]{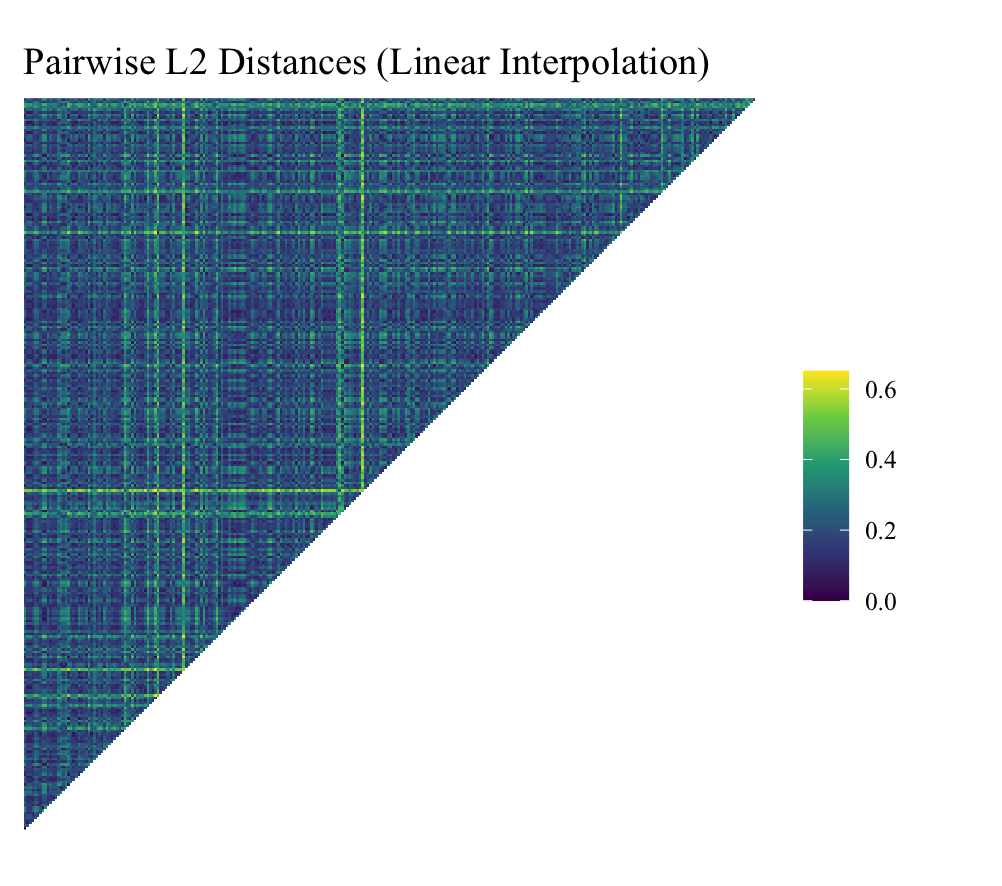}\\
    \caption{Mean pairwise distance matrices estimated using \texttt{LFCM} for interpolation/extrapolation across 500 generated sample paths (left), and linear interpolation/extrapolation for the same week (right).}
    \label{fig:mcmc-heatmaps}
\end{figure}

The left panel of Figure \ref{fig:mcmc-heatmaps} shows the mean pairwise distance matrices for 293 devices estimated on 500 generated sample paths from individual/week \texttt{LFCM} models. The right panel shows the same matrix after linear interpolation.
There is relative agreement between both interpolation/extrapolation strategies, which reveal significant structure in the distance matrices, holding valuable implications for social and behavioral science. The clear vertical and horizontal lines suggest that some individuals remain consistently far from others - potentially indicating social isolation or distinct spatial domains - while small dark patches point to local clusters,  which may reflect subgroups exhibiting confined or periodic movement. Note that the \texttt{LFCM}-based distance matrix has relatively fewer outliers, which may be due to the poor extrapolation behavior of the linear method. This spatial patterning may offer critical insights to understand community structure, mobility, and the design of targeted interventions and policies.

\section{Discussion} \label{sec:discussion}

We have shown that the L\'evy flight cluster model (\texttt{LFCM}) robustly captures the intricate dynamics of human mobility, as evidenced by both simulation studies and real-world GPS analyses. Our proposed model successfully delineates probabilistic activity spaces by distinguishing between localized activity clusters characterized by stationary repetitive behavior and sporadic long-range movements consistent with Lévy flight dynamics. These dual components were reliably estimated despite irregular sampling, and the posterior distributions aligned well with established mobility metrics such as jump length, radius of gyration, and mean square displacement. In addition, the ability of the model to generate probabilistic sociomatrices via interpolation and extrapolation further underscores its efficacy in representing temporal and spatial overlap among individuals. This capacity is pivotal for examining phenomena such as social isolation, geographic mobility constraints, and the emergence of localized subgroups, thus offering nuanced insights into patterns of social exclusion and differential access to resources.

Overall, the \texttt{LFCM} provides a substantial methodological advancement over traditional approaches by accurately modeling human mobility through a generative Bayesian framework that integrates both local Brownian motion and long-range L\'{e}vy jumps. The success of the model in reproducing individual-level activity distributions, while simultaneously offering uncertainty bounds, highlights its potential as a tool for both rigorous scientific inquiry and practical applications in urban planning, epidemiology, and data anonymization. By allowing researchers to simulate realistic activity spaces and derive probabilistic overlap measures even under irregular sampling conditions, the \texttt{LFCM} lays a comprehensive foundation for future investigations into the behavioral and social dimensions of human mobility.

\section*{Acknowledgment}

AHW and GSC thank the following university departments for their in-kind support through their affiliate faculty positions: Department of Statistical Sciences and Operations Research at the Virginia Commonwealth University (AHW, GSC), Department of Statistics at the University of Washington (GSC), and the Department of Statistics and Actuarial Science at the University of Waterloo (GSC).


\appendix

\section{Collapsed MCMC sampler for the LFCM}\label{appendix:collapsedmcmc}

The \texttt{LFCM} can be ``collapsed'' through integrating out nuisance parameters, significantly lessening computational complexity. Below we derive the collapsed MCMC sampling algorithm for the \texttt{LFCM}.  Let  $n_g = \sum_{i=1}^n\ \mathbf{1}\{c_i = g\}$, $n_b = \sum_{i=1}^n b_i$, $n_\eta = \sum_{i=1}^n \eta_i$, $\mathfrak{Z} = \{(g,l)\ \mid\ g = 1,\ldots, N_G;\ \ l = 1,\ldots, L_g\},$ and $n_z = \sum_{i=1}^n \mathbf{1}\{z_i = (g,l)\}$. Let $\mathbf{\Delta X}$ represent $n$ observed spatial movements and let $\mathfrak{H}$ represent the collection of $n\times 1$ latent vectors $\underline{\mathfrak{h}}$. The joint distribution of the data and model parameters can be written as
\begin{equation}
    \begin{aligned}
        f(\mathbf{\underline{\Delta X}}, \mathfrak{H},\underline{\theta},N_G,\underline{\Delta t}) &= f(\mathbf{\underline{\Delta X}}, \mathfrak{H},\underline{\omega},\nu,\alpha,\eta,\underline{\xi},\{\underline{\mu}_g,\mathbf{\Sigma}_g^*\}_{1:G},N_G,\underline{\Delta t}).
    \end{aligned}
\end{equation}

If one is only interested in the latent vectors and hyperparameters associated with the Brownian motion, the corresponding marginal joint distribution is:
\begin{equation}
    \begin{gathered}
        \nonumber
        \int f(\mathbf{\Delta X}, \mathfrak{H}, \underline{\theta}, N_G,  \underline{\Delta t})\ d\underline{\omega}\ d\nu\ d\alpha\ d\eta\ d\underline{\xi} \propto\\
         \phantom{\propto}\underbrace{\dfrac{\prod_{g=1}^{N_G} \Gamma(a_g + n_g)}{\Gamma(N + \sum a_g)}}_{\text{activity group probability}}
        \phantom{\propto}\times \underbrace{\dfrac{\Gamma(\beta_1 + n_b)\Gamma(\beta_2 + n - n_b)}{\Gamma(\beta_1 + \beta_2 + n)}}_{\text{jump probability}}
        \phantom{\propto}\times \underbrace{\dfrac{\Gamma(\upvarpi + n_b)}{(\upzeta + \ln\frac{\|\Delta x\|_2}{\varepsilon})^{\upvarpi + n_b}}}_{\text{conditional jump distribution}}\times\\
        \phantom{\propto}\underbrace{\dfrac{\Gamma(\mathpzc{b}_1 + n_\eta)\Gamma(\mathpzc{b}_2 + n_b - n_\eta)}{\Gamma(\mathpzc{b}_1 + \mathpzc{b}_2 + n_b)}}_{\text{return probability}}
        \phantom{\propto}\times\underbrace{\dfrac{\prod_{g=1}^{N_K} \Gamma(\mathpzc{z}_k + n_k)}{\Gamma(N_K + \sum \mathpzc{z}_k)}}_{\text{return group probability}}.
    \end{gathered}
\end{equation}
The hyperparameters associated with the Brownian motion can also be integrated out of the marginal joint distribution \--- Appendix \ref{appendix:derivation}:
\begin{equation}
    \begin{aligned}
        \nonumber
        \int f(\mathbf{\Delta X}, \mathfrak{H}, \underline{\theta}, N_G,  \underline{\Delta t})\ d\underline{\mu}_gd\mathbf{\Sigma}_g^* & \propto 2^{\nu_{\tildelow{n}_g} + 1}(2\pi)^{\tilde{n}_g} |\mathbf{K}_{\tildelow{n}_g}^{-1}|^{-(\nu_{\tildelow{n}_g} + 1)/2}
        \times\Gamma_2\left(\frac{\nu_{\tildelow{n}_g} + 1}{2}\right),
    \end{aligned}
    \label{eq:post_ci}
\end{equation}
where $\mathbf{K}_{\tildelow{n}_g} \equiv (\kappa\mathbf{I} + n\overline{\mathbf{B}}^\top\overline{\mathbf{B}})$, $\nu_{\tildelow{n}_{g}} \equiv \nu + \tilde{n}_g$, and
\begin{align*}
    \mathbf{W}_{\tildelow{n}_g}^{-1} &\equiv \sum_{i=1}^{\tildelow{n}_g} (\mathbf{A}_i^{1/2}\underline{x}_i - \overline{\mathbf{A}^{1/2}\underline{x}})(\mathbf{A}_i^{1/2}\underline{x}_i - \overline{\mathbf{A}^{1/2}\underline{x}})^\top - \\
    &\phantom{\equiv}\tilde{n}_g\kappa\overline{\mathbf{A}^{1/2}\underline{x}}(\kappa\mathbf{I} + \tilde{n}_g\overline{\mathbf{B}}^{\top}\overline{\mathbf{B}})^{-1}\overline{\mathbf{A}^{1/2}\underline{x}}^\top + \mathbf{W}^{-1}.
\end{align*}
Here $\tilde{n}_g = \# \{\underline{\Delta x}_i : (c_i, b_i) = (g, 0)\text{ or }(b_i,\eta_i, z_i) =  (1,0,(g,l))\}$ is the number of spatial movements that are classified as a Brownian motion movement or a return, and the matrices $\mathbf{A},\mathbf{B},\mathbf{D}$ are given by $\mathbf{B}_i = \mathbf{A}_i^{1/2}\mathbf{D}_i$,
\begin{subequations}
    \begin{equation*}
        \begin{aligned}
            \nonumber
            \mathbf{D}_i &\equiv 
            \begin{cases}
                \Delta t_i \mathbf{I} & (c_i, b_i) = (g,0),\\
                \frac{1}{2}T_z\mathbf{I} & (b_i, \eta_i, z_i) = (1,0,(g,l)),
            \end{cases}
        \end{aligned}
    \end{equation*}
    \begin{equation*}
        \begin{aligned}
            \nonumber
            \mathbf{A}_i^{-1} &\equiv 
            \begin{cases}
                \Delta t_i \mathbf{I} & (c_i, b_i) = (g, 0),\\
                \frac{1}{3}T_z\mathbf{I} & (b_i, \eta_i, z_i) = (1, 0, (g,l)).
            \end{cases}
        \end{aligned}
    \end{equation*}
\end{subequations}
The marginal joint distribution becomes
\begin{equation}
    \begin{aligned}
        f(\Delta \mathbf{X}, \mathfrak{H}, N_G, \underline{\Delta t}) &\propto
        f_{z}(\underline{z} \mid \underline{b}, \underline{c}, \underline{\eta}, \mathbf{\Delta X}, N_G) \times 
        f_{\Delta x}(\Delta \mathbf{X} \mid \mathfrak{H}, N_G, \underline{t})\times \\
         &\quad f_{b, \eta}(\underline{b}, \underline{\eta} \mid \underline{c}, N_G)
        \times f_c(\underline{c}\ \mid\ N_G) \times \pi(N_G),
    \end{aligned}
    \label{eq:full_marginal}
\end{equation}
with $\pi(N_G)$ the marginal distribution on the number of groups $N_G$,

\begin{eqnarray*}
 f_c(\underline{c} \mid N_G) &=& \quad \dfrac{\prod_{g=1}^{N_G} \Gamma(a_g + n_g)}{\Gamma(n + \sum a_g)} \dfrac{\Gamma(\sum_g a_g)}{\prod_g \Gamma(a_g)},\\
f_{b}(\underline{b} \mid \underline{c}) &=&
            \dfrac{\Gamma(\beta_1 + n_b)\Gamma(\beta_2 +n - n_b)}{\Gamma(\beta_1 + \beta_2 + n)}\dfrac{\Gamma(\beta_1 + \beta_2)}{\Gamma(\beta_1)\Gamma(\beta_2)},\\
f_{\eta}(\underline{\eta} \mid \underline{b}) &=&
            \dfrac{\Gamma(\mathpzc{b}_1 + n_\eta)\Gamma(\mathpzc{b}_2 + n_b - n_\eta)}{\Gamma(\mathpzc{b}_1 + \mathpzc{b}_2 + n_b)}\dfrac{\Gamma(\mathpzc{b}_1 + \mathpzc{b}_2)}{\Gamma(\mathpzc{b}_1)\Gamma(\mathpzc{b}_2)},\\
f_{z}(\underline{z} \mid \underline{b}, \underline{c}, \underline{\eta}, N_G) &=& \dfrac{\prod_{z=1}^{N_z} \Gamma(\xi_z + n_z)}{\Gamma(n_{\eta} + \sum \xi_z)} \dfrac{\Gamma(\sum_z \xi_z)}{\prod_z \Gamma(\xi_z)},
\end{eqnarray*}
and
\begin{eqnarray*}
        f_{\Delta x}(\Delta \mathbf{X}\ \mid \mathfrak{H}, N_G, \Delta\underline{t}) &= &\prod_{g=1}^{N_G} \biggl[\dfrac{2^{\nu_{\tildelow{n}_g} + 1}(2\pi)^{\tildelow{n}_g}|\mathbf{K}_{\tildelow{n}_g}|^{1/2}|\mathbf{W}_{\tildelow{n}_g}^{-1}|^{-\frac{\nu_{\tildelow{n}_g} + 1}{2}}}{2^{\nu + 1}2\pi\kappa|\mathbf{W}^{-1}|^{-\frac{\nu + 1}{2}}}
         \frac{\Gamma_2\left(\frac{\nu_{\tildelow{n}_g} + 1}{2}\right)}{\Gamma_2\left(\frac{\nu + 1}{2}\right)}\biggr],\\
        &=& \quad \dfrac{2^{2n - 1}\pi^{n - 1}\gamma^{\nu + 1}}{\kappa\Gamma_2\left(\dfrac{\nu + 1}{2}\right)}
         \prod_{g=1}^{N_G} \Gamma_2\left(\dfrac{\nu_{\tildelow{n}_g} + 1}{2}\right)  |\mathbf{K}_{\tildelow{n}_g}|^{1/2}|\mathbf{W}_{\tildelow{n}_g}^{-1}|^{-\frac{\nu_{\tildelow{n}_g} + 1}{2}},\\
        &=& \quad \prod_{g=1}^{N_G} f^{(g)}(\Delta \mathbf{X}, \mathfrak{H},  N_G, \Delta \underline{t}).
\end{eqnarray*}
Hence, the joint distribution $f(\Delta \mathbf{X}, \mathfrak{H}, N_G, \Delta \underline{t})$ is discrete in the latent vectors and the number of groups---$(\underline{c}, \underline{b}, \underline{z}, \underline{\eta}, N_G)$---and stochastic searches over the discrete space of finite mixtures can be used to sample from the posteriors without reversible jumps \citep{ryan2017bayesian, richardson1997bayesian}. This discrete search technique has been successful in a number of finite mixture applications \citep[e.g.,][]{ryan2017bayesian, wyse2012block, nobile2007bayesian}.  The exact probability of the proposed movement can be calculated as
\begin{equation}
    \begin{aligned}
        \frac{f(\Delta \mathbf{X}, \mathfrak{H}, N_G, \underline{\Delta t})}{\sum f(\Delta \mathbf{X}, \mathfrak{H}', N_G, \underline{\Delta t})},
    \end{aligned}
    \label{eq:renormalize}
\end{equation}
and often reduces to computationally simpler forms through cancelation of the non-varying terms from Eq. \eqref{eq:full_marginal}, depending on the update being considered.

\paragraph*{Conditional distribution of $\eta$}The conditional distribution of $\eta_i$ is constructed as a Beta-Bernoulli($\mathpzc{p, b_1, b_2}$). Hence,

\begin{eqnarray*}
        f_{\Delta J^*}(\underline{\Delta x}; \vartheta, \alpha) &=& \left((1-\mathfrak{p})\left[\xi_{z}\mathcal{N}(\underline{x}(t_i)\ ; \underline{\widetilde{\mu}}_{z}, \mathbf{\Sigma}_{z})\right]^{\mathbf{1}\{z = (g,l)\}}\right)^{\mathbf{1}\{\eta=0\}}\mathpzc{p}^{\mathbf{1}\{\eta=1\}}.    
\end{eqnarray*}

\vspace{.1in}

\paragraph*{Conditional distribution of $N_G$} Conditional sampling of the number of activity groups $N_G$ is performed using the absorption / ejection method of \citet{ryan2017bayesian}. Let $p^{\text{ej.}} = p^{\text{ab.}} = 0.5$. If $N_G > 1$ and $N_G < M_G$ for some maximum number of groups $M_G$, then with probability $p^{\text{ej.}}$ a ``eject'' move is made, and with probability $p^{\text{ab.}}$ an ``absorb'' move is made. If $N_G = 1$, the eject move is made with probability $1$, and if $N_G = M_G$, the absorb move is made with probability $1$. 

The eject move chooses one of the $N_G$ clusters at random and attempts to reallocate a random number of the $L_g$ continuous paths in $g$. The number of paths with probability $\varrho \sim \text{Beta}(\mathpzc{a},\mathpzc{a})$ is selected to move from $g$ to a new cluster, $N_G + 1$. The proposal, denoted below by the superscript $`*'$, is accepted with probability:
\vspace{.05in}
\begin{eqnarray*}
    1 \vee \dfrac{f^{(g)}(\mathbf{\Delta X}, \mathfrak{H}^*, \underline{\Delta t})f^{(N_G + 1)}(\mathbf{\Delta X}, \mathfrak{H}^*, \underline{\Delta t})}{f^{(g)}(\mathbf{\Delta X}, \mathfrak{H}, \underline{\Delta t})}\frac{\pi(N_G + 1)}{\pi(N_G)}\dfrac{\Gamma(\mathpzc{a})^2}{\Gamma(2\mathpzc{a})}\dfrac{\Gamma(2\mathpzc{a} + n_g)}{\Gamma(\mathpzc{a} + n_g^*)\Gamma(\mathpzc{a} + n_{N_G + 1}^*)},
\end{eqnarray*}
\vspace{.05in}
where $\pi(N_G)$ represents the marginal distribution on the number of groups$N_G$.

The absorb move is performed by randomly selecting two components $g_1, g_2$ from the $N_G + 1$ groups and combining them into a single component. This move is accepted with probability:
\vspace{.05in}
\begin{eqnarray*}
        1 \vee \dfrac{f^{(g_1)}(\mathbf{\Delta X}, \mathfrak{H}^*, \underline{\Delta t})}{f^{(g_1)}(\mathbf{\Delta X}, \mathfrak{H}, \underline{\Delta t})f^{(g_2)}(\mathbf{\Delta X}, \mathfrak{H}, \underline{z}, \underline{\Delta t})}\dfrac{\pi(N_G)}{\pi(N_G+1)}\dfrac{\Gamma(2\mathpzc{a})}{\Gamma(\mathpzc{a})^2}\dfrac{\Gamma(\mathpzc{a} + n_{g_1})\Gamma(\mathpzc{a} + n_{g_2})}{\Gamma(2\mathpzc{a} + n_{g_1}^*)}.    
\end{eqnarray*}
\vspace{.1in}
\paragraph*{MCMC Sampling}Sampling of the marginal posterior distribution 
$$
 f(\mathfrak{H}, N_G, \underline{\Delta t}\mid \Delta \mathbf{X}) \propto f(\Delta \mathbf{X}, \mathfrak{H}, N_G, \underline{\Delta t})
$$
in Eq. \eqref{eq:full_marginal} is conducted using a Markov chain Monte Carlo (MCMC) sampling algorithm. A single sweep of parameter updates (see Table \ref{table:model_params} in the main text) are described in \texttt{Algorithm \ref{algo:gibbs}} below. We start with two types of movement data. First, we have the spatial difference data, which tells us how far an individual has moved. Second, we have the data for the time difference, indicating the time it took for these movements to take place. We are interested in grouping movements into what we call ``activity groups'' or categories of movement such as walking, running, or jumping. We also want to understand specific characteristics of these movements, for example, whether they correspond to a trip into a new spatial region or whether they represent a return trip to a spatial region that has been previously visited.

We start with an initial assignment of movements into groups and their classifications (e.g., exploratory vs. return movements) that is generated by sampling from a hypergeometric distribution. Next we sequentially cycle through the following steps:\\

\begin{itemize}
\item[1.] \textbf{Update the group classifications.} For each group, we look at every movement and decide if its current group assignment is good or if it needs to be updated. This decision is based on the spatial and temporal data and the characteristics of the movements.
\item[2.] \textbf{Adjust the number of groups.} We consider if we need more or fewer groups to accurately represent the variety of movements.
\item[3.] \textbf{Refine the movement characteristics.} For each movement, we revisit our initial guesses about its characteristics (like whether it's an exploration or a return) and update these guesses to better match the data.
\item[4.] \textbf{Classify the activity areas.} We decide if each movement is towards a new area (exploration) or a known area (return). This helps in understanding how individuals or objects interact with their environment over time.\\
\end{itemize}

With each repetition of these steps, the movement groupings and their classification are refined. This algorithm is a systematic way to sift through data on movements, categorizing them into different types of activities, and understanding their nature.

\begin{algorithm}[htbp]
\caption{\texttt{LFCM} (MCMC)}\label{algo:gibbs}
{\small
\hspace*{\algorithmicindent} \textbf{Input} Individual spatial difference data $\mathbf{\Delta X}$, time difference data $\underline{\Delta t}$,\\
\hspace*{\algorithmicindent} hyperparameters (see Table \ref{table:model_params} in main text).\\
\hspace*{\algorithmicindent} \textbf{Require} Maximum \# activity groups $M_G$, minimum jump length $\varepsilon$\\
\hspace*{\algorithmicindent} \textbf{Initialize} $\underline{b} =\text{HyperGeometric}(n,M_G-1,n)$, 
$\underline{c} = \text{Multin}(n,N_G^0,(N_G^0)^{-1}\underline{1})$,\\
\hspace*{\algorithmicindent} $\underline{\eta} = \text{Bern}(n,1/2) \odot \underline{b}$, $\underline{z} = f(\underline{c},\underline{b})$.\\
\hspace*{\algorithmicindent} $L_g = f(\underline{z})$ \# contiguous points in activity group.
\begin{algorithmic}
\For{$g = 1,\ldots, N_G$} \algorithmiccomment{Update activity group classification}
    \For{$l = 1,\ldots, L_g$}
        \If{$b_i = 0$}
            \State Sample $\underline{c}_{l} \propto f_c(\underline{c}_l\ |\ \underline{c}_{-l}, \underline{b}, \underline{\eta}, \underline{z}, \Delta \mathbf{X}, \underline{\Delta t})$
        \EndIf
    \EndFor
\EndFor
\State
\State Propose $N_G \leftarrow N_G \pm 1$ \algorithmiccomment{Absorption/Ejection Step}
\State
\For{$i = 1,\ldots, N$}
\algorithmiccomment{Update jump classification}
        \State Sample $b_i \propto f_b(b_i\ |\ \underline{b}_{-i}, \underline{c}, \underline{\eta}, \underline{z}, \Delta \mathbf{X}, \underline{\Delta t})$
\EndFor
\State
\For{$i = 1,\ldots, N$}  \algorithmiccomment{Update Exploration/Return classification}
    \State Sample $\eta_i \propto  f_{\eta}(\eta_i\ |\ \underline{\eta}_{-i}, \underline{c}, \underline{b}, \underline{z}, \Delta \mathbf{X}, \underline{\Delta t})$ 
\EndFor
\State
\For{$i = 1,\ldots, N$}  \algorithmiccomment{Update Activity Area classification}
    \If{$\eta_i = 0$}
            \State Sample $z_i \propto f_{z}(z_i\ |\ \underline{z}_{-i}, \underline{c}, \underline{b}, \underline{\eta}, \mathbf{\Delta X}, \underline{\Delta t})$
    \EndIf
\EndFor
\end{algorithmic}
}
\caption{A single MCMC sweep for the \texttt{LFCM}.}
\end{algorithm}

\section{Derivations of posterior distributions} \label{appendix:derivation}

For positive definite symmetric matrices $\mathbf{D},\ \mathbf{A}$, define the prior normal and Wishart distributions
\begin{eqnarray*}
    \underline{x}_i &\sim& \mathcal{N}\left(\mathbf{D}_i\underline{\mu},\ \left(\mathbf{A}_i^{1/2}\mathbf{\Lambda}\mathbf{A}_i^{1/2}\right)^{-1} \right),\\
    \underline{\mu} &\sim& \mathcal{N}\left(\underline{0}, \left(\kappa\mathbf{\Lambda}\right)^{-1} \right),\\
    \mathbf{\Lambda} &\sim& \mathcal{W}\left(\nu, \mathbf{W}\right).\\
\end{eqnarray*}
The corresponding densities are proportional to
\begin{eqnarray*}
    &|\mathbf{\Lambda}|^{1/2}\prod_{i} |\mathbf{A}_i|^{1/2}
    \exp\left[
    -\frac{1}{2} (\underline{x}_i - \mathbf{D}_i\underline{\mu})^\top \mathbf{A}_i^{1/2}\mathbf{\Lambda}\mathbf{A}_i^{1/2}(\underline{x}_i - \mathbf{D}_i\underline{\mu})\right], & \label{eq:a1}\\
    &|\kappa\mathbf{\Lambda}|^{1/2}\exp\left[-\frac{\kappa}{2}
    \underline{\mu}^\top\mathbf{\Lambda}\underline{\mu}
    \right],& \label{eq:a2}\\
    & |\mathbf{\Lambda}|^{(\nu - 2)/2}\exp\left[-\frac{1}{2}\text{tr}(\mathbf{W}^{-1}\mathbf{\Lambda})\right].& \label{eq:a3}
\end{eqnarray*}

The exponent of Eq. \eqref{eq:a1} can be written as

\begin{align*}
    -\frac{1}{2}((\mathbf{A}_i^{1/2}\mathbf{D}_i(\mathbf{D}_i^{-1}\underline{x}_i - \underline{\mu}))^\top\mathbf{\Lambda}(\mathbf{A}_i^{1/2}\mathbf{D}_i(\mathbf{D}_i^{-1}\underline{x}_i - \underline{\mu})).
\end{align*}

Letting $\mathbf{B}_i := \mathbf{A}_i^{1/2}\mathbf{D}_i$, the quadratic expression above is expanded to

\begin{align*}
    \sum_{i=1}^N \underline{x}_i^{\top}\mathbf{A}_i^{1/2}\mathbf{\Lambda}\mathbf{A}_i^{1/2}\underline{x}_i - N\overline{\mathbf{A}^{1/2}\underline{x}}^{\top}\mathbf{\Lambda}\overline{\mathbf{B}}\underline{\mu} - N\underline{\mu}^\top\overline{\mathbf{B}}^\top\mathbf{\Lambda}\overline{\mathbf{A}^{1/2}\underline{x}} +
    N\underline{\mu}^{\top}\overline{\mathbf{B}}^\top\mathbf{\Lambda}\overline{\mathbf{B}}\underline{\mu},
\end{align*}

\noindent where the overscore of a matrix, e.g. $\overline{\mathbf{A}}$, represents the mean operator. The term in Eq. \eqref{eq:a3} can be combined with the exponent of Eq. \eqref{eq:a2} to become

\begin{align*}
    \underline{\mu}^{\top}\overline{\mathbf{B}}^{\top}(\kappa\overline{\mathbf{B}}^{-\top}\mathbf{\Lambda}\overline{\mathbf{B}}^{-1} + N\mathbf{\Lambda})\overline{\mathbf{B}}\underline{\mu} 
    &= \underline{\mu}^{\top}(\kappa\mathbf{\Lambda} + N\overline{\mathbf{B}}^{\top}\mathbf{\Lambda}\overline{\mathbf{B}})\underline{\mu}.
\end{align*}

Since $\overline{\mathbf{B}}\mathbf{\Lambda} = \mathbf{\Lambda}\overline{\mathbf{B}}$ when $\mathbf{B}_i = \mathbf{A}_i^{1/2}\mathbf{D}_i$, this becomes

$$\underline{\mu}^{\top}(\kappa\mathbf{I} + N\overline{\mathbf{B}}^{\top}\overline{\mathbf{B}})\mathbf{\Lambda}\underline{\mu}.$$

Adding and subtracting the term

\begin{align*}
    N\overline{\mathbf{B}}^{\top}\overline{\mathbf{B}}(\kappa{\mathbf{I}} + N\overline{\mathbf{B}}^\top\overline{\mathbf{B}})^{-1}\overline{\mathbf{A}^{1/2}\underline{x}}^\top\mathbf{\Lambda}\overline{\mathbf{A}^{1/2}\underline{x}}
\end{align*}

allows for the factorization

\begin{align*}
    (\kappa{\mathbf{I}} + N\overline{\mathbf{B}}^\top\overline{\mathbf{B}})\left(\underline{\mu} - N\overline{\mathbf{B}}^{\top}\overline{\mathbf{B}}(\kappa{\mathbf{I}} + N\overline{\mathbf{B}}^\top\overline{\mathbf{B}})^{-1}\overline{\mathbf{A}^{1/2}\underline{x}} \right)^\top \mathbf{\Lambda}\left(\underline{\mu} - N\overline{\mathbf{B}}^{\top}\overline{\mathbf{B}}(\kappa{\mathbf{I}} + N\overline{\mathbf{B}}^\top\overline{\mathbf{B}})^{-1}\overline{\mathbf{A}^{1/2}\underline{x}} \right).
\end{align*}

Adding and subtracting the term $N\overline{\mathbf{A}^{1/2}\underline{x}}^\top\mathbf{\Lambda}\overline{\mathbf{A}^{1/2}\underline{x}}$ allows factorization 

\begin{align*}
    \sum_{i=1}^N (\mathbf{A}_i^{1/2}\underline{x}_i - \overline{\mathbf{A}^{1/2}\underline{x}})^\top\mathbf{\Lambda}(\mathbf{A}_i^{1/2}\underline{x}_i - \overline{\mathbf{A}^{1/2}\underline{x}}).
\end{align*}

Combining the other two terms:

\begin{align*}
    &N\overline{\mathbf{A}^{1/2}\underline{x}}^\top\mathbf{\Lambda}\overline{\mathbf{A}^{1/2}\underline{x}} -N^2\overline{\mathbf{B}}^{\top}\overline{\mathbf{B}}(\kappa\mathbf{I} + N\overline{\mathbf{B}}^\top\overline{\mathbf{B}})^{-1}\overline{\mathbf{A}^{1/2}\underline{x}}^\top\mathbf{\Lambda}\overline{\mathbf{A}^{1/2}\underline{x}}\\
    &=
    N\kappa(\kappa\mathbf{I} + N\overline{\mathbf{B}}^\top\overline{\mathbf{B}})^{-1}
    \overline{\mathbf{A}^{1/2}\underline{x}}^\top\mathbf{\Lambda}\overline{\mathbf{A}^{1/2}\underline{x}}.
\end{align*}

This is a scalar and is equal to its own trace. Moreover, since the trace has a cyclic commutative property, we have

\begin{align*}
   N\kappa(\kappa\mathbf{I} + N\overline{\mathbf{B}}^\top\overline{\mathbf{B}})^{-1}
    \overline{\mathbf{A}^{1/2}\underline{x}}^\top\mathbf{\Lambda}\overline{\mathbf{A}^{1/2}\underline{x}} &= \text{tr}\left( N\kappa\overline{\mathbf{A}^{1/2}\underline{x}}(\kappa\mathbf{I} + N\overline{\mathbf{B}}^{\top}\overline{\mathbf{B}})^{-1}\overline{\mathbf{A}^{1/2}\underline{x}}^\top\mathbf{\Lambda}\right)
\end{align*}

and

\begin{align*}
    \sum_{i=1}^N (\mathbf{A}_i^{1/2}\underline{x}_i - \overline{\mathbf{A}^{1/2}\underline{x}})^\top\mathbf{\Lambda}(\mathbf{A}_i^{1/2}\underline{x}_i - \overline{\mathbf{A}^{1/2}\underline{x}}) &= \text{tr}\left( \sum_{i=1}^N (\mathbf{A}_i^{1/2}\underline{x}_i - \overline{\mathbf{A}^{1/2}\underline{x}})(\mathbf{A}_i^{1/2}\underline{x}_i - \overline{\mathbf{A}^{1/2}\underline{x}})^\top\mathbf{\Lambda}\right).
\end{align*}

These can be combined with Eq. \eqref{eq:a3} so that the exponent becomes

\begin{align*}
    \text{tr}\left(\left(\sum_{i=1}^N (\mathbf{A}_i^{1/2}\underline{x}_i - \overline{\mathbf{A}^{1/2}\underline{x}})(\mathbf{A}_i^{1/2}\underline{x}_i - \overline{\mathbf{A}^{1/2}\underline{x}})^\top - N\kappa\overline{\mathbf{A}^{1/2}\underline{x}}(\kappa\mathbf{I} + N\overline{\mathbf{B}}^{\top}\overline{\mathbf{B}})^{-1}\overline{\mathbf{A}^{1/2}\underline{x}}^\top + \mathbf{W}^{-1}\right)\mathbf{\Lambda}\right).
\end{align*}

The posterior parameters are then

\begin{align*}
    \mathbf{K}_n &= (\kappa\mathbf{I} + N\overline{\mathbf{B}}^\top\overline{\mathbf{B}}), \\
    \underline{\mu}_n &= N\overline{\mathbf{B}}^{\top}\overline{\mathbf{B}}(\kappa\mathbf{I} + N\overline{\mathbf{B}}^{\top}\overline{\mathbf{B}})^{-1}\overline{\mathbf{A}^{1/2}\underline{x}},\\
    \nu_n &= \nu + N,\\
    \mathbf{W}_n^{-1} &= \sum_{i=1}^N (\mathbf{A}_i^{1/2}\underline{x}_i - \overline{\mathbf{A}^{1/2}\underline{x}})(\mathbf{A}_i^{1/2}\underline{x}_i - \overline{\mathbf{A}^{1/2}\underline{x}})^\top - N\kappa\overline{\mathbf{A}^{1/2}\underline{x}}(\kappa\mathbf{I} + N\overline{\mathbf{B}}^{\top}\overline{\mathbf{B}})^{-1}\overline{\mathbf{A}^{1/2}\underline{x}}^\top + \mathbf{W}^{-1}.
\end{align*}

When $\mathbf{A} = \mathbf{D} = \mathbf{I}$, the posterior distribution becomes the standard posterior parameters of a Normal-Wishart distribution. The normalizing constant is 

\begin{align*}
    2^{\nu_n + 1}(2\pi)^{N}|\mathbf{K}_n|^{1/2}|\mathbf{W}_n^{-1}|^{-(\nu_n + 1)/2}\Gamma_2\left(\frac{\nu_n + 1}{2}\right).
\end{align*}

For $\{\underline{\Delta x}_i\ :\ c_i = g\}$, 
$$\mathbf{D}_i = \mathbf{A}_i^{-1} = \Delta t_i\mathbf{I}.$$ For $\{\underline{\Delta x}_i\ :\ b_i = 1, \eta_i = 0, z_i = (g,l)\}$, 
$$\mathbf{D}_i = \frac{1}{2}T^{(g,l)}\mathbf{I},\qquad \mathbf{A}_i^{-1} = \frac{1}{3}T^{(g,l)}\mathbf{I}.$$

\section{Similarity metrics for activity areas} \label{appendix:similaritymetrics}

Several similarity metrics can be applied in the context of evaluating the convergence of estimated activity areas. In this section we discuss the Kullback-Leibler (KL) divergence, the Hausdorff distance, the Jaccard distance, and finally, a modification of the latter called the Overlap.

The {Kullback-Leibler (KL) divergence} provides a measure of the difference between two probability distributions. Let \(P\) and \(Q\) be two probability distributions defined on the same measurable space.
The measure quantifies how much information is lost when the distribution \(P\) is approximated by \(Q\).
For discrete and continuous variables, respectively, it is defined as
\[
D_{\mathrm{KL}}(P \,\|\, Q) \;=\;
\begin{cases}
\displaystyle
\sum_{x} P(x)\,\log\frac{P(x)}{Q(x)} & \text{(discrete)},\\[1.1em]
\displaystyle
\int P(x)\,\log\frac{P(x)}{Q(x)}\,dx & \text{(continuous)}.
\end{cases}
\]
 Its formula is essentially an expectation of the logarithmic difference between the probabilities \( P \) and \( Q \), where \( P \) and \( Q \) are the two distributions being compared. This metric is appropriate when ``activity areas'' are conceptualized as probability distributions. However, its sensitivity is notably pronounced when one distribution assigns a significant probability to an event that the other largely disregards, leading to potential large divergences even for small differences. As seen in Figure \ref{act_over_time}, the number of activity areas naturally increases over time, leading to difficulty in interpreting this metric.

The {Hausdorff distance} gauges the maximum distance between two point sets, encapsulating the idea of shape dissimilarity. It captures the greatest distance from a point in one set to the closest point in the other set. This metric is inherently sensitive to outliers, as even well-aligned datasets can yield a high Hausdorff distance if a single point pair exhibits a large distance. Due to the sparsity of the data, activity areas in early weeks are likely to have far greater spread than those in later weeks.

The {Jaccard distance}, derived from the Jaccard index, evaluates the dissimilarity between two discrete sets. Defined as 
\[ D_J(A, B) = 1 - \frac{|A \cap B|}{|A \cup B|}, \]
this metric quantifies the proportion of non-overlapping elements in two sets. It is particularly suitable when ``activity areas'' are represented as discrete locations. Given its foundational basis on ratios, the Jaccard distance offers a balanced sensitivity to changes, especially in large datasets where individual differences might have proportionally smaller impacts.

Finally, we propose the Overlap distance \(1 - \frac{|A \cap B|}{|B|}\) which computes the proportion of elements in set \(B\) that are absent from set \(A\), serving as an asymmetric dissimilarity measure. Contrasting this with the symmetric Jaccard distance, which considers mutual overlaps, the asymmetric measure emphasizes the extent to which \(A\) covers \(B\). Specifically, it captures the dissimilarity value when observing \(A\) relative to \(B\), focusing on the coverage of \(B\) by \(A\). In applications such as evaluating ``activity areas'', this measure provides insight into the degree of coverage of the activity areas estimated with 12 weeks of data by the activity areas estimated with a fewer number of weeks, offering a practical perspective of the coverage of ``true'' activity areas, with less sensitivity than the Jaccard distance.

\end{document}